\documentclass[acmsmall]{acmart}

\AtBeginDocument{%
  }

\setcopyright{cc}
\setcctype{by}
\acmJournal{PACMHCI}
\acmYear{2026} \acmVolume{10} \acmNumber{6} \acmArticle{CSCW060}
\acmMonth{10} \acmDOI{10.1145/3816908}

\usepackage{subcaption}

\begin{document}

\title{Causal Language in Post Titles Shapes Deeper Topological Structures of Online Conversations}



\author{Zhuoyu Shi}

\affiliation{%
  \institution{Thomas Lord Department of Computer Science, University of Southern California}
  \city{Los Angeles}
  \state{California}
  \country{United States}
}

\affiliation{%
  \institution{Information Sciences Institute, University of Southern California}
  \city{Los Angeles}
  \state{California}
  \country{United States}
}
\email{zhuoyush@usc.edu}

\author{Fred Morstatter}

\affiliation{%
  \institution{Thomas Lord Department of Computer Science, University of Southern California}
  \city{Los Angeles}
  \state{California}
  \country{United States}
}

\affiliation{%
  \institution{Information Sciences Institute, University of Southern California}
  \city{Los Angeles}
  \state{California}
  \country{United States}
}

\email{morstatt@usc.edu}


\begin{abstract}
Causal reasoning is fundamental to human understanding and information organization. People prefer causal explanations because they offer coherence, predictability, and a sense of control.  Conversational structures shape how knowledge and perspectives are shared, validated, and amplified in networked publics. Understanding the structural effects of causal language can reveal pathways to fostering deeper, more meaningful interactions online. In this work, we investigate how causal language influences the topology and temporal evolution of discussion threads in online conversations with a dataset of 17 million posts across 200 subreddits in 2023 on Reddit. Our results show that causal language is consistently associated with deeper, more sustained conversations, with effects emerging early in the lifecycle of a thread, as demonstrated through a counterfactual experiment. Importantly, emotional responses do not differ substantially between causal language and non-causal language, suggesting that structural depth arises from framing itself rather than affective escalation. A lightweight qualitative analysis shows that causal framing titles prompt users to elaborate more with reasoning and contribute personal experiences, supporting deeper multi-turn exchanges. These findings suggest that causal language acts not merely as a stylistic device, but as a cognitively grounded and structurally influential signal that shapes the topological structures of online conversations. 
\end{abstract}

\ccsdesc[500]{Human-centered computing~Empirical studies in collaborative and social computing}

\keywords{Online behavior, Social Media, Online Discussion, Reddit, Causal Language}

\received{May 13 2025}
\received[revised]{January 13 2026}
\received[accepted]{April 9 2026}

\maketitle

\section{Introduction}

In complex systems, small differences in initial conditions can lead to vastly different long-term outcomes. This principle, central to nonlinear dynamics and complexity science, has been widely observed in natural and social systems alike \cite{strogatz2024nonlinear, mitchell2009complexity}. In the context of online communication, the linguistic framing of a post may serve as such an initial condition, subtly guiding the trajectory of discussion that follows.

\begin{figure}[tbhp]
\centering
\includegraphics[width=.99\linewidth]{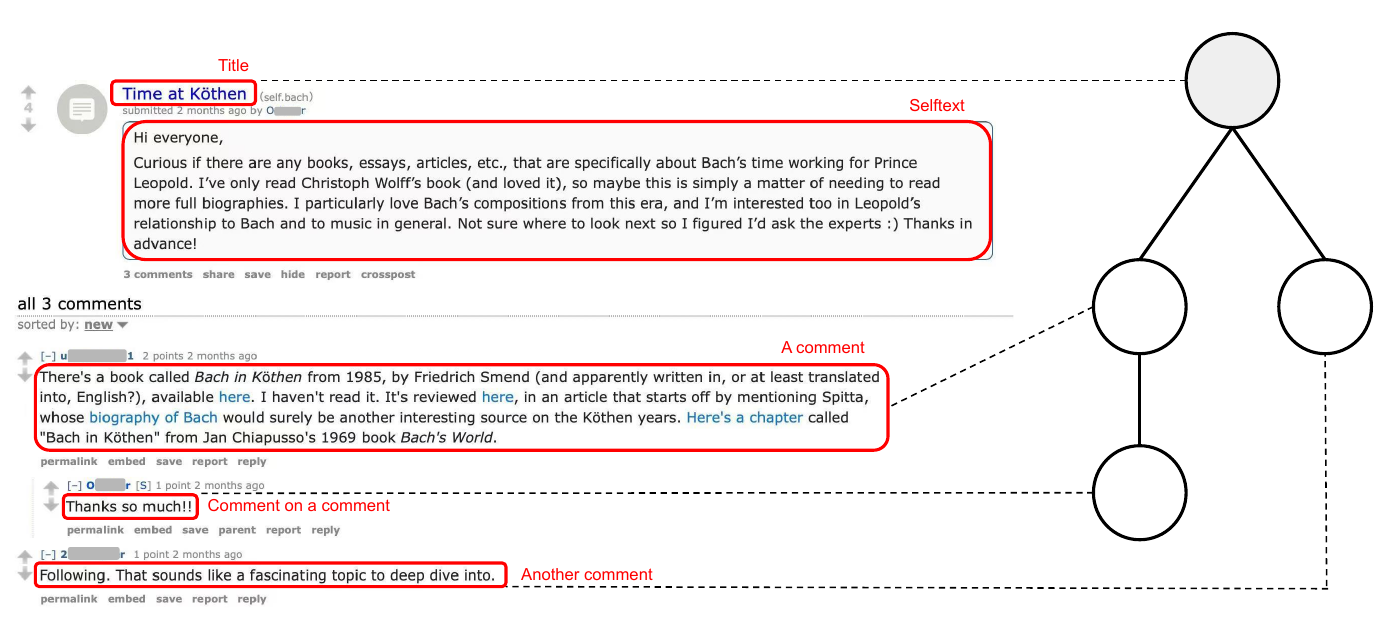}
\caption{Illustration of the tree structure of a Reddit post and its comment tree.}
\label{fig:tree_illustration}
\end{figure}

Causal language plays a central role in how people interpret, explain, and engage with information. Prior work in psychology and communication has shown that humans are especially attuned to cause-effect structures in language, often preferring causal explanations over purely descriptive ones because they offer a sense of coherence, predictability, and control \cite{lombrozo2006structure, gopnik2004theory}. In social contexts, causal framing can influence attribution, emotional response, and behavioral intent \cite{weiner1985attributional, lagnado2008judgments, langer1978mindlessness}.

 Prior research on online discussion evolvement has highlighted structural patterns in conversation threads and the influence of early user activity on long-term engagement \cite{gomez2008statistical, kumar2010dynamics, medvedev2019modelling}. These studies often model discussions as branching or networked processes shaped by platform affordances, user behavior, and timing \cite{aragon2017generative, medvedev2017anatomy, yu2024characterizing}. Structural engagement is not just a descriptive property of conversations but a driver of community sustainability. Prior work shows that threaded, multi-reply interactions increase user retention and long-term participation by enabling users to return, be responded to, and become embedded in interactional loops \cite{budak2017threading, gomez2013likelihood, krohn2019modelling}. Deep and branching conversations also create opportunities for identity formation, norm reinforcement, and collective sense-making that shallow exchanges cannot sustain \cite{kumar2010dynamics, medvedev2019modelling}.  On digital platforms, where attention is scarce and interaction is rapid, even brief uses of causal language, such as ``because,'' ``leads to,'' or ``results in'', could shape how users perceive a post’s purpose, relevance, or argumentative stance. Understanding the structural effects of causal language can help   foster deeper, more meaningful interactions online. This suggests a need to more closely examine how causal language itself, beyond topic or sentiment, guides the structure and dynamics of social media discourse.

This study investigates whether causal language in post titles acts as a structural signal that influences the trajectory and complexity of ensuing online discussions. Social media platforms like Reddit provide a valuable context for this inquiry: each post initiates a public conversation tree, and title phrasing is often the first content users encounter before choosing to engage. Our goal is to understand whether the presence of causal language at the outset corresponds with differences in the volume, structure, temporal evolution, and emotional tone of user replies.

To answer this, we analyze a comprehensive dataset of all posts and comments in 200 Reddit communities (subreddits) over a full calendar year. This enables us to track how discussions evolve from the moment of posting and to test whether differences in discussion structure correlate with the presence or absence of causal framing. Importantly, we go beyond simple correlational analysis by incorporating a counterfactual matching design to evaluate whether causal language itself and not just confounders like topic, timing, or emotion is associated with differences in conversational structure. To this end,  we address several significant research questions, namely:

\begin{itemize}
    
    \item  \textbf{RQ1} - Do causal and non-causal language spark different discussion topologies?
    
    \item  \textbf{RQ2} - How do discussions sparked by causal and non-causal language differ in their topological evolution?

    \item \textbf{RQ3} Are structural differences in discussion driven by early emotional responses to causal language?

    \item \textbf{RQ4} Does the presence of causal language influence the emotion of discussions?

    \item \textbf{RQ5} Are structural differences in discussion driven by explanation-reasoning and experience sharing triggered by causal language?

\end{itemize}

To address these questions, we begin by detecting the presence of causal language in post titles using a state-of-the-art deep learning model. We apply this model to 17 million Reddit posts across 200 subreddits in 2023, to identify titles with or without causal language. Then, we analyze whether posts with causal language in their titles produce different structural patterns in their comment threads compared to posts without causal language in their titles, using a counterfactual setting to control for key confounds. Subsequently, we examine the temporal dynamics of discussion growth, analyzing how comment trees evolve over time and whether causal framing influences the pacing and trajectory of engagement. We also investigate the emotion of comments, both at the early stages of discussion and across entire discussion trees, to test whether causal language triggers greater emotional intensity, which might explain deeper structural outcomes. Last, we conduct a lightweight qualitative analysis of early reply-to-reply interactions to help explain the underlying mechanisms.

Our findings show that causal language is consistently associated with deeper, more sustained, and structurally complex conversations. These effects emerge early in the discussion and persist over time, even after controlling for potential confounds. The influence of causal framing emerges early in the lifecycle of a conversation. Notably, emotional responses do not differ significantly between causal and non-causal groups, suggesting that the structural divergence is not driven by affective escalation but by the framing given by causal language itself. Qualitative evidence further suggests that causal titles prompt users to elaborate more with reasoning and contribute personal experiences, supporting deeper multi-turn exchanges. Together, these findings offer new insights into how causal language shapes discussion patterns of interaction in online discourse, suggesting that causal framing could help platform designers foster richer, more sustained conversations rather than focusing solely on surface-level engagement. More broadly, they highlight how causal framing shapes collective sense-making and knowledge exchange in networked publics.

\section{Related Work}

Previous work on online discussions modeled the structural growth of conversation threads as branching processes. For example, studies found that thread trees exhibit heavy-tailed distributions in depth and breadth, suggesting patterns consistent with self-reinforcing reply behavior dynamics and reinforcement mechanisms in user replies \cite{gomez2008statistical, kumar2010dynamics}. Researchers have showed that user interactions tend to follow predictable cascades conditioned on early activity \cite{gomez2013likelihood, medvedev2019modelling, medvedev2017anatomy}. Aragón et al. \cite{aragon2017generative} provided a comprehensive review of such generative models. Additional work examined the relationship between structural characteristics (e.g., reply tree width, depth, and temporal pacing) and the design of the platform itself, such as the introduction of threaded replies \cite{aragon2017thread}, which significantly changed user engagement and branching behavior. Budak et al. \cite{budak2017threading} further demonstrated that even minimal threading features can increase user retention, particularly among new commenters, suggesting structural affordances play a role in shaping participation.

\begin{figure}[tbhp]
\centering
\includegraphics[width=.95\linewidth]{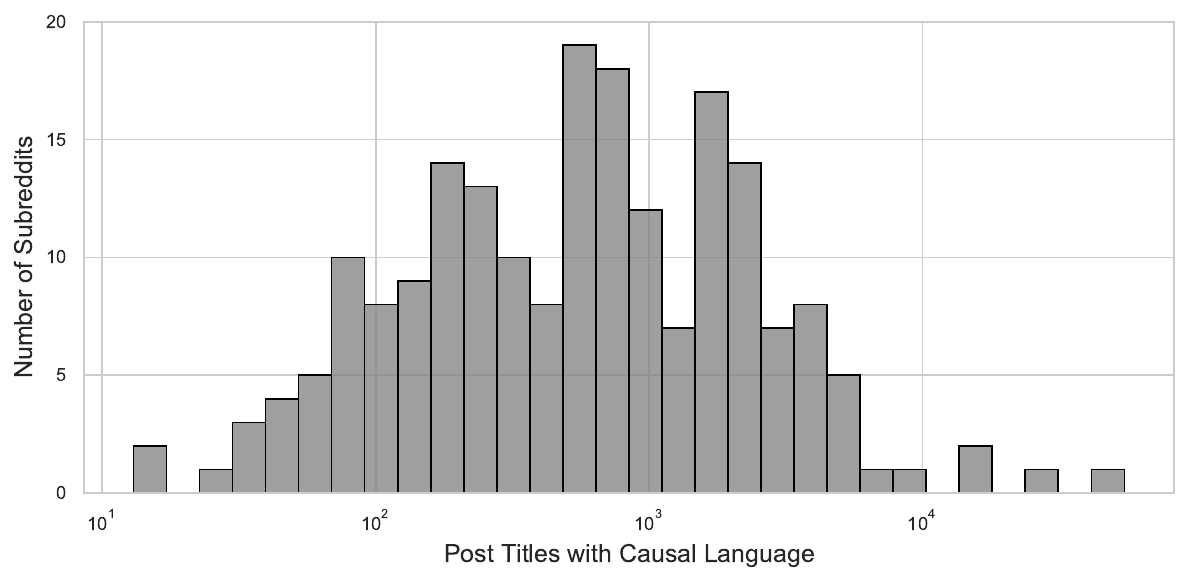}
\caption{The distribution of posts containing causal language across all 200 subreddits.}
\label{fig:subreddit_CL_dist}
\end{figure}

Some studies expanded on this by modeling Reddit threads as interaction networks, capturing not only the topology but also dynamic properties such as growth speed and resilience to external stimuli. For instance, Medvedev et al. \cite{medvedev2017anatomy} and Goglia and Vega \cite{goglia2024structure} treated discussions as evolving user graphs, revealing structural differences between subreddits and contrasting Reddit with other online forums. Yu et al. \cite{yu2024characterizing} quantified how global and local post attributes affect thread shape, finding that local signals more strongly influence reply tree topology than global features. Other predictive models focused on the size or lifespan of conversations. Krohn and Weninger \cite{krohn2019modelling} used submission text to forecast a thread’s ultimate scope, while Horawalavithana et al. \cite{horawalavithana2022online} predicted discussion growth using content and timing features. Additional research has characterized the role of individual users in sparking or sustaining engagement; Saveski et al. \cite{saveski2021social} identified “social catalysts” who prompt interaction among others, while Zhang et al. \cite{zhang2018characterizing} described distinct patterns of participant interactions that shape discussion trajectories. Platform governance has also been shown to influence thread dynamics, where Chandrasekharan et al. \cite{chandrasekharan2017you} demonstrated that subreddit bans reduced hate speech without significant spillover into other communities. Still, most of this work ignores how causal language in the post title might bias the resulting thread structure. Our paper builds on this line of inquiry by introducing content-specific causal framing as a measurable upstream factor in structural thread outcomes.

Our work bridges these literatures by treating causal language in post titles as an explicit initial condition that can shape the topology of ensuing conversations. A large body of research in psychology and cognitive science suggests that humans have a deep-seated drive to understand the world in causal terms. Even infants seek out cause-effect relationships, using interventions and observations to build intuitive causal maps of their environment \cite{gopnik2004theory,gopnik2007causal}. This tendency continues through development, as children and adults alike organize knowledge around causal schemas, not merely associations \cite{keil2006explanation,sloman2009causal}. Causal explanations are particularly satisfying, often preferred over statistically superior but non-causal alternatives, because they confer a sense of understanding and control \cite{lombrozo2006structure}. The theory-theory framework posits that people construct and revise mental models of how the world works driven by causal coherence and domain-specific expectations \cite{gopnik2012reconstructing}. These mental models scaffold our sense of narrative, prediction, and interpretation, suggesting that causal language is not merely a linguistic flourish but a reflection of deeper cognitive structures.

Causal language is central to how people process stories and social events. Narrative comprehension research has long shown that readers mentally represent stories in terms of cause-effect chains: events with stronger causal links are remembered better and judged as more important \cite{graesser1994constructing}. Discourse studies further demonstrate that causal connectives like ``because'' and ``so'' form a key class of coherence relations, guiding interpretation and making discourse comprehensible \cite{sanders1992toward}. In social reasoning, causal attributions influence not only memory but emotion, judgment, and behavior: people’s beliefs about the causes of success, failure, or social outcomes shape responses ranging from blame to policy preferences \cite{weiner1985attributional,iyengar1987television}. The form of causal explanation therefore serves as a potent framing device in how information is understood and evaluated.

In communication, causal framing has profound consequences for belief and engagement. Studies show that how events are causally framed can shift blame, emotion, and behavior \cite{lagnado2008judgments}. Even minimal use of causal language (e.g., ``because'') increases compliance and persuasiveness, highlighting listeners’ expectations for cause-effect structure in discourse \cite{langer1978mindlessness}. From an epistemological perspective, causal stories, far more than lists of facts, organize public understanding, whether in law, journalism, or everyday conversation \cite{pennington1992explaining}. Causal language thus functions not only to explain but to persuade, structure attention, and signal relevance in dialogue.

Despite extensive work in cognitive and social sciences, causal language remains understudied in online discourse, where it may be especially consequential. Social media environments reward brief, high-impact posts, conditions ripe for simplified causal framing. Previous study suggests that causal language shapes how information spreads in online social networks \cite{shi2024diffusion}, yet few studies have linked such framing to the structure of ensuing conversations. This represents a key gap: if causal language activates explanation-seeking or reasoning, it may encourage deeper or more contentious discussions. In structured forums like Reddit, where discussions unfold as tree-like reply graphs, such language features could affect how people interact. Bridging causal cognition with conversational topology opens a new avenue for understanding how ideas propagate and evolve, not just through networks of people, but through structures of response and conversation.

\section{Data Description}

To investigate how causal language in post titles influences the structural dynamics of online discussions, we leverage a comprehensive dataset from Reddit~\cite{redditdata}. This dataset spans every publicly available Reddit post and comment from June 2005 to December 2023. In our study, we focus on the most active and popular communities by selecting the top 200 subreddits as ranked in Reddit's official \textit{Best of Reddit}\footnote{\url{https://www.reddit.com/best/communities/1/?rdt=52192}} list. To ensure recency and consistency in user behavior and platform design, we restrict our analysis to a single calendar year: January to December 2023 (inclusive). Within this period, we observe a total of 17,196,132 posts across the selected 200 subreddits. To understand text-based discourse and minimize confounding effects introduced by external content (e.g., images, videos, or hyperlinks), we filter out posts that include URLs in their selftext. This results in a refined dataset of 16,811,872 post titles, with associated 343,865,806 user comments. This filtered dataset provides a robust foundation for analyzing the impact of causal language on the structural, temporal, and emotional dimensions of online conversations.

\begin{table}

  \begin{tabular}{p{0.52\linewidth} p{0.44\linewidth}}
\toprule

 \multicolumn{1}{c}{\centering Titles with Causal Language} & \multicolumn{1}{c}{\centering Titles without Causal Language} \\

\midrule

TIL grain-free dog food is not healthier for our dogs as it can \textbf{lead to} heart disease.  & Who Can Relate? \\
\addlinespace[2.5pt]

work is \textbf{causing} me anxiety, how do you deal?  & I am going to try to grow wasabi. \\
\addlinespace[2.5pt]
 
Terrible forearm/elbow pain \textbf{from} pull ups  & Where do I go next? \\
\addlinespace[2.5pt]
 
Every concealer I try \textbf{gives} me pimples  & looking for Job \\
\addlinespace[2.5pt]
 
How day trading \textbf{made} me poor  & Is it possible to install a case fan? I don’t see any screw holes… \\
    
    \bottomrule
  \end{tabular}

  \caption{Examples of titles with or without causal language. Causal connections are in bold. Causal language refers to phrases, sentences, or discourse that express cause-and-effect relationships between different elements. For example, in the first example, ``TIL grain-free dog food is not healthier for our dogs as it can \textbf{lead to} heart disease.'' contains causal language, where there is a cause ``grain-free dog food'' and an effect ``heart disease,'' connected by a causal connection ``\textbf{lead to}.'' }
  \label{tab:CL_text}
\end{table}

\section{Detecting Causal Language in Post Titles}

To identify the presence of causal language in Reddit post titles, we employ a state-of-the-art deep learning pipeline~\cite{priniski2023pipeline}. This model builds on RoBERTa~\cite{liu2019roberta}, a transformer-based language model, and has been fine-tuned specifically for causal relation detection. Beyond binary classification, the model identifies spans corresponding to the cause and effect within a sentence, enabling fine-grained semantic analysis. The model achieves high performance, with an F1-score of 0.874, a precision of 0.883, and a recall of 0.865~\cite{priniski2023pipeline}. We apply this classifier to the titles of all posts in our dataset.
Out of 16,811,872 eligible posts, 330,225 contain causal language in the title, while the rest do not. Figure~\ref{fig:subreddit_CL_dist} illustrates the distribution of posts containing causal language across all 200 subreddits. 

To ensure sufficient representation across communities for comparative analysis, we exclude subreddits with fewer than 200 posts that contain causal language. 
This filtering process naturally removes subreddits which are largely visual and contain minimal textual discussion (e.g., \textit{r/gifs} and \textit{r/photoshopbattles}).
After filtering, 144 subreddits remain. Within this refined set, we observe 324,161 posts with causal language in the title and 15,751,294 without (see Table~\ref{tab:CL_text} for examples).

\section{Validation of Causal Language Detection}

To validate our detection results, we hire annotators to label a total of 100 messages, employing a random sampling strategy to select these messages. Specifically, 50 messages were randomly selected from each of the causal language group, and the non-causal language group. Annotators were hired from Prolific. Each annotator was trained with four examples with explanations, and then tested with six straightforward examples. We only include annotators who correctly annotated at least five out of six of these examples. Each annotator was tasked with annotating 25 messages and compensated \$4.5 for their effort. Each message was labeled as ``contains CL'' or  ``does not contain CL'' by three distinct annotators to ensure reliability. Final labels were determined via majority vote to reduce annotator bias. For an example of an annotation, please refer to Figure~\ref{fig:anno_CL} in the Appendix.

The precision achieved by the CL group is 84\%, while the non-CL group attains a precision of 92\%. To further evaluate the consistency among annotators, we compute inter-annotator reliability using Krippendorff’s alpha. The resulting alpha is $\alpha = 0.41$ for the CL group and  $\alpha = 0.70$ for the non-CL group. The overall agreement is found to be 70.67\% for the CL group, and 85.33\% for the non-CL group. Importantly, despite moderate agreement in the CL group, the classifier achieves high precision, indicating strong and consistent alignment with majority human judgments. This provides evidence that annotator disagreement is not random noise, but instead reflects structured ambiguity concentrated in borderline cases where causal relationships are less apparent to the outlier annotator. In such cases, individual annotators may reasonably differ, yet the majority label remains stable across annotator sets. The high precision therefore demonstrates that the signal captured by the model corresponds closely to the dominant human interpretation, even in the presence of localized disagreement. These results indicate robust performance, sufficient for large-scale identification of causal language across Reddit communities.

\section{RQ1. Structural Differences in Conversation Trees}

Our first research question examines whether causal language in post titles is associated with different structural patterns in the following discussions. Specifically, we seek to understand whether posts framed with causal language spark conversation trees that differ in size, depth, and width compared to those without such framing. These topological metrics allow us to characterize not only the volume of participation but also the structural depth and distribution of engagement, which are key indicators of how discourse unfolds in online settings.

The volume of posts, discussion dynamics, and thematic focus vary widely across subreddits. Some communities are highly active and discussion-oriented, while others are more content-driven. Additionally, the frequency of causal language in post titles differs substantially across subreddits, as shown in Figure~\ref{fig:subreddit_CL_dist}. In addition, other confoundings, such as topic, timing, and emotional framing, may influence discussion structure. To control for these confounding factors and ensure a balanced comparison, we adopt a counterfactual matching design. Specifically, we construct a treatment group consisting of posts with causal language and a control group of posts without causal language. To isolate the effect of causal language as much as possible, each post in the treatment group is matched to a non-causal counterpart that is as similar as possible across a set of pre-defined covariates~\cite{stuart2010matching}. This results in matched pairs of posts that are comparable in the key respects that matter,  except for the presence of causal framing in the title. This design enables us to make stronger claims about the potential effect of language framing on discussion structure. For all posts across all 144 subreddits without a counterfactual design, please see Figure~\ref{fig:RQ1_all_144_all} in the Appendix.

\subsection{Selecting Covariates for Similarity Measurement}

To create high-quality matched pairs between posts with and without causal language, we identify a set of covariates that reflect key contextual, emotional, and semantic aspects of a post. Specifically, we match on four covariates: subreddit, time, emotion, and semantic content, as prior literature shows their influence on engagement, content dynamics, and discourse structure in online settings \cite{brady2017emotion, stieglitz2013emotions, krohn2022subreddit, oddny2023impact, razis2020modeling}.

\begin{itemize}
    \item \textbf{Subreddit}: Posts from the same subreddit are likely to share norms, audience expectations, and topical focus. Matching within the same subreddit controls for community-level variation.

    \item \textbf{Time}: Posts are matched within the same calendar month. This accounts for temporal variation such as changes in platform activity, real-world events, and seasonal or topical trends.

    \item \textbf{Emotion}: Emotional content influences both user attention and the likelihood of engagement. Informed by the Basic Emotion Theory \cite{ekman1999basic}, we adopt a framework built around six core emotions: fear, anger, joy, sadness, disgust, and surprise. Each post is categorized into one of these six emotions or into a seventh ``neutral'' category. These foundational emotions offer a psychologically grounded way to represent affective tone, which plays a critical role in shaping interpersonal communication and cognitive framing \cite{berger2012makes, brady2017emotion}. We classify each post’s emotional tone using a state-of-the-art language model-based emotion classifier \cite{hartmann2022emotionenglish}. Each post is categorized into one of these six emotions: fear, anger, joy, sadness, disgust, and surprise, or into a seventh neutral category.

    \item \textbf{Semantic Content}: Posts are matched on meaning using sentence embeddings derived from Sentence-BERT \cite{reimers-2019-sentence-bert}, which generates dense vector representations capturing the semantic structure of each title.
\end{itemize}

\subsection{Matching Method}

We adopt a pairwise matching procedure guided by best practices in causal inference using observational data \cite{stuart2010matching}. Our goal is to create pairs of posts that are similar across all selected covariates but differ in one critical aspect: the presence or absence of causal language in the title.

For each post in the treatment group (i.e., titles with causal language), we perform the following steps. From the pool of titles without causal language, we filter for candidates that exactly match the treatment post in three categorical covariates: subreddit, calendar month, and emotion category. Among the eligible candidates, we compute the cosine similarity between sentence embeddings generated with Sentence-BERT of the treatment title and each non-causal candidate. We select the non-causal title with the highest cosine similarity to serve as the control match for the treatment title.

This procedure ensures that both the treatment and control groups are balanced across structural, emotional, temporal, and semantic dimensions. The key distinguishing factor between the two groups is the presence of causal language, which allows for a cleaner estimation of its effect on discussion structure.

\subsection{Measurement of Structural Differences}
For each post, we construct its associated comment tree, where the root node represents the post title, and each subsequent node represents a user comment. Directed edges indicate reply relationships between comments, forming a tree-structured conversation graph. An illustration of this structure is provided in Figure~\ref{fig:tree_illustration}.

To quantify the topological structures of these discussions, we compute the following metrics:

\begin{itemize}
    
    \item  Size: The total number of comments in the post, including the root post. This captures the overall volume of participation.
    
    \item  Width: The maximum number of sibling comments at any depth level in the tree. This captures how broad the conversation is at its most active layer.
    
    \item  Depth: The length of the longest comment chain from the title to the most deeply nested comment. This reflects how sustained the interaction becomes.

    \item Average Depth: The average length of all comment chains from the title to each leaf comment in the tree. This reflects the typical level of sustained interaction in the discussion.

\end{itemize}

\subsection{Matching Results and Validation}

Our matching procedure yields 324,161 high-quality matched pairs (see Table~\ref{tab:CF_pairs} for illustrative examples), which means that all posts with titles containing causal language are matched to a post with a title without causal language. The average semantic cosine similarity within each pair was M = 0.9228 (SD = 0.0327), indicating that the two titles (with vs. without causal language) in each pair are highly similar at the semantic level.

We first create a baseline group, where we randomly select 200 posts with causal language and 200 posts without causal language from each of the 144 subreddits that met our inclusion criteria described in section 4. This results in a balanced sample of 57,600 posts. We then take all the 28,800 posts with causal language from the baseline group (200 random posts from each subreddit), and their counterparts matched in this section without causal language, results in a balanced sample of 57,600 posts.

To validate the reliability of our counterfactual process, we conducted a human evaluation study. We randomly sampled 50 title pairs (with and without causal language) from our dataset after the counterfactual experiment. For each title with causal language, we also paired it with a title without causal language from the baseline group. Thus, each title with causal language was paired with two titles without causal language: one from the baseline group and one from the counterfactual experiment. We then ask human annotators to determine which title without causal language is semantically closer to the corresponding title with causal language.

Human annotations were collected via the Prolific platform. Annotators were evaluated on three test cases, and we only included those who correctly annotated all three. Each qualifying annotator evaluated 25 comparisons and received a compensation of \$2.50.

Each comparison was independently labeled by three annotators. Final labels were determined by majority vote to ensure consistency and reduce individual annotator bias. Figure~\ref{fig:anno_CF} provides an example of the annotation interface.

\begin{table}
  \small
  \begin{tabular}{p{1.5cm} p{0.42\linewidth} p{0.42\linewidth}}
    \toprule

\multicolumn{1}{c}{\centering}{} & \multicolumn{1}{c}{\centering Treatment Group} & \multicolumn{1}{c}{\centering Control Group} \\

\midrule

Example 1 & The oven turned off \textbf{due to} power flickering  & Food processor woes \\
\addlinespace[2.5pt]

Example 2 & Subtle joint pain weeks \textbf{after} exercising & Back pain even when my lower back is on the ground?  \\
\addlinespace[2.5pt]

Example 3 & Expensive products \textbf{made} my hair worse. (CW: ED) & Rant about hair stylist sibling  \\

    \bottomrule
  \end{tabular}
  \caption{Examples of pairs of titles with and without causal language via counterfactual analysis. Causal connections are in bold. }
  \label{tab:CF_pairs}
\end{table}

Results show that 92\% of the titles without causal language in the counterfactual experiment were judged to be semantically closer to their causal counterparts, while only 8\% of the titles without causal language from the baseline group were judged closer. To evaluate consistency among annotators, we calculated inter-annotator agreement using Fleiss’ Kappa. The overall agreement was 0.88, indicating high reliability. These results suggest that our counterfactual experiment is effective and substantially suggests the strength of the counterfactual experiment.

\subsection{Structural Results}
To understand how causal language in post titles relates to the structural dynamics of conversation trees, we analyze the distributions of key topological metrics: size, depth, and width across posts with and without causal framing. Rather than common measures such as means or standard deviations, which can be highly sensitive to extreme values, we use percentile-based analysis to provide a more stable and interpretable view of the data. Social media data often exhibits long-tailed distributions, with a small number of posts generating disproportionately large amounts of engagement. These high-variance outliers can obscure general trends and inflate summary statistics. By focusing on percentiles, we mitigate the influence of such outliers and gain a clearer picture of typical structural patterns within each group. Specifically, we examine the metric distributions up to the 99th percentile to ensure that our comparisons reflect meaningful differences in user engagement without being dominated by rare, anomalously large threads.

\begin{figure}[tbhp]
\centering

\begin{subfigure}[b]{0.48\linewidth}
    \centering
    \includegraphics[width=\linewidth]{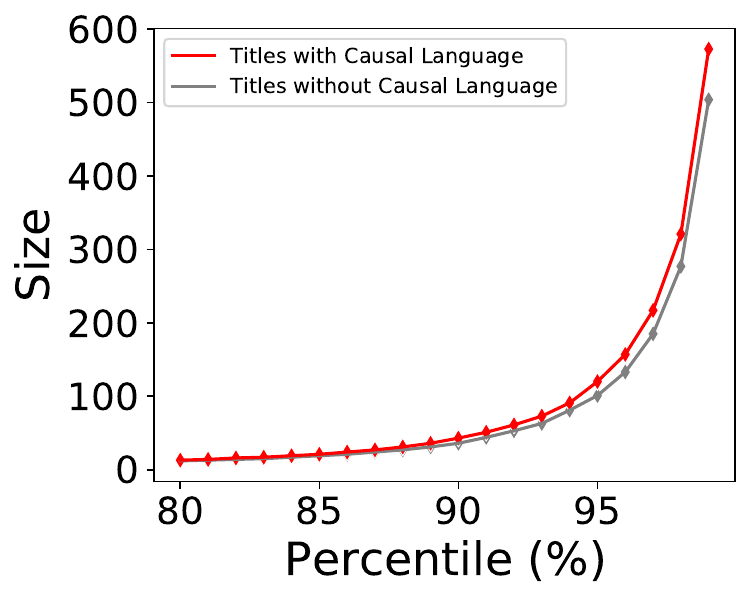}
    \caption{Comparison of post title \textbf{size} distributions between causal and non-causal language via counterfactual analysis. The 99th percentile of causal titles reaching 573, compared to 504 for non-causal titles.}
    \label{fig:RQ2_CF_144_200_size}
\end{subfigure}
\hfill
\begin{subfigure}[b]{0.48\linewidth}
    \centering
    \includegraphics[width=\linewidth]{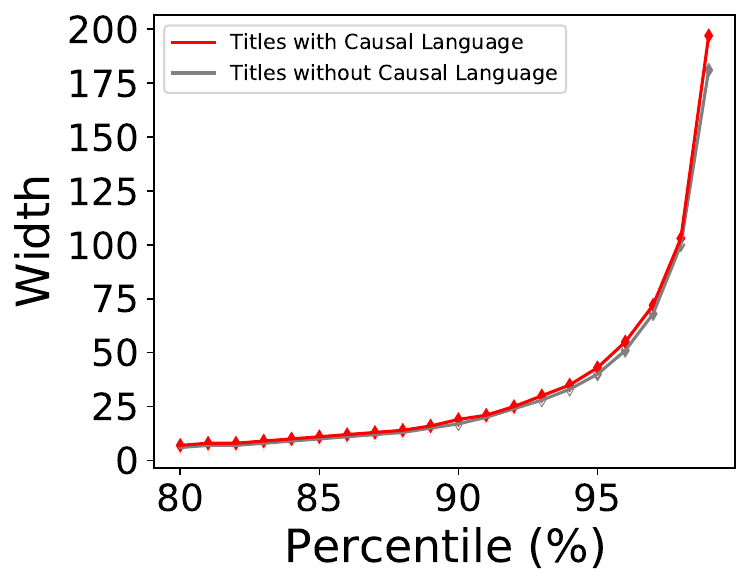}
    \caption{Comparison of post title \textbf{width} distributions between causal and non-causal language via counterfactual analysis. At the 99th percentile, causal titles reach a width of 197, while non-causal titles is 181.}
    \label{fig:RQ2_CF_144_200_width}
\end{subfigure}

\vspace{1em} 

\begin{subfigure}[b]{0.48\linewidth}
    \centering
    \includegraphics[width=\linewidth]{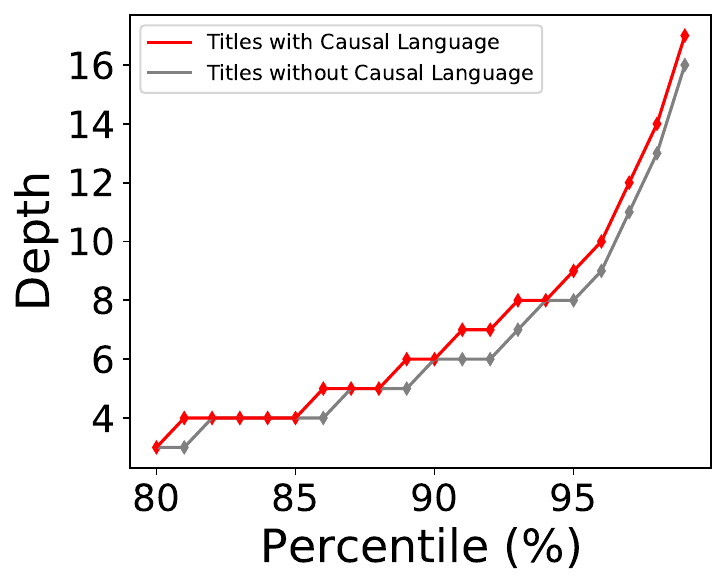}
    \caption{Comparison of post title \textbf{depth} distributions between causal and non-causal language via counterfactual analysis. At the 99th percentile, causal titles reach a depth of 17, while non-causal titles remain at 16.}
    \label{fig:RQ2_CF_144_200_depth}
\end{subfigure}
\hfill
\begin{subfigure}[b]{0.48\linewidth}
    \centering
    \includegraphics[width=\linewidth]{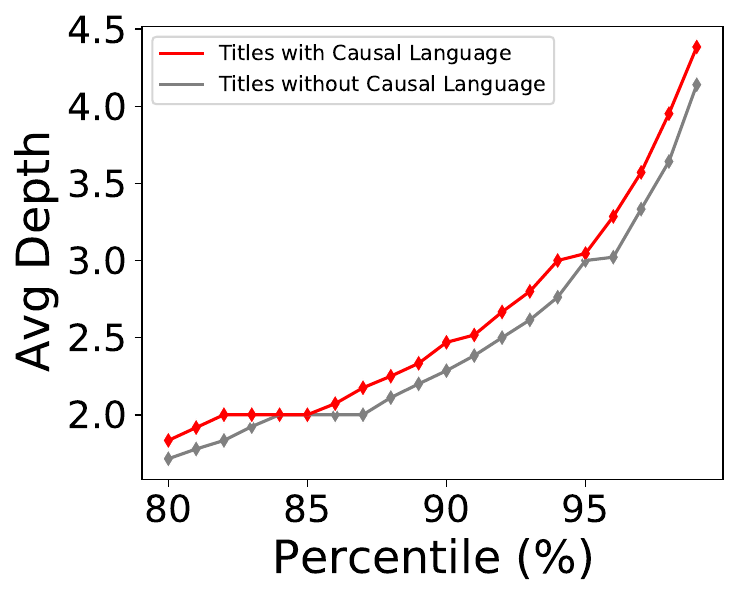}
    \caption{Comparison of post title \textbf{average depth} distributions between causal and non-causal language via counterfactual analysis. At the 99th percentile, causal titles reach an average depth of 4.38, while non-causal titles remain at 4.14.}
    \label{fig:RQ2_CF_144_200_avg_depth}
\end{subfigure}

\caption{Comparison of paired post title \textbf{size}, \textbf{width}, \textbf{depth}, and \textbf{average depth}} distributions between causal and non-causal language after counterfactual matching.
\label{fig:RQ2_CF_144_200}
\end{figure}

As shown in Figure~\ref{fig:RQ2_CF_144_200_size}, posts with causal language in the title are consistently associated with larger conversation trees across the distribution. By the 99th percentile, causal posts attract 573 comments, compared to 504 for non-causal posts, a difference of approximately 100 comments. This suggests that causal framing significantly co-occurs with more participation in the discussion, particularly in highly active threads. The consistently higher size across percentiles indicates that this effect is not limited to extreme outliers but generalizes across a wide range of engagement levels. The Kolmogorov–Smirnov test reveals a significant difference ($p$-value $=5.42 \times 10^{-8}$). Although the effect size is modest (D = 0.025), the result indicates that the shift in distribution, while small, is systematic and robust.

Figure~\ref{fig:RQ2_CF_144_200_width} presents the width distribution, defined as the greatest number of parallel comments at any level in the tree. While the pattern is similar to size in that causal posts consistently show higher values, the magnitude of difference is less pronounced. At the 99th percentile, causal posts reach a width of 197, compared to 181 for non-causal posts. This suggests that causal framing may modestly co-occur with more horizontal branching. However, the effect on width is relatively subtle, indicating that causal language primarily goes deeper (see Figure~\ref{fig:RQ2_CF_144_200_depth} and and~\ref{fig:RQ2_CF_144_200_avg_depth}), rather than broader, engagement. K-S test suggests a significant difference ($p$-value $= 2.09 \times 10^{-7}$). The effect size is small (D = 0.024), yet the consistency of the finding demonstrates that the distributional change is real and not attributable to random variation.

Figure~\ref{fig:RQ2_CF_144_200_depth} suggests notable structural difference at the maximum depth of comment chains. At the 99th percentile, causal posts reach a depth of 17, while non-causal posts stay at 16. This indicates that posts with causal framing titles have deeper multi-turn interactions, where such structural depth is often associated with deeper conversational engagement. The K-S test confirms a significant difference between the two groups ($p$-value $=5.82 \times 10^{-7}$). Despite the modest effect size (D = 0.023), this distributional shift, though subtle, is both systematic and robust.

Figure~\ref{fig:RQ2_CF_144_200_avg_depth} reveals a clear structural difference in average comment-chain depth. At the 99th percentile, causal posts exhibit an average depth of 4.38, compared to 4.14 for non-causal posts. This pattern indicates that titles with causal framing are associated with more sustained multi-turn exchanges, a characteristic commonly linked to deeper conversational engagement. The K-S test confirms that the two distributions differ significantly ($p$-value $=7.95 \times 10^{-9}$). Although the effect size remains small (D = 0.026), the consistency of the shift suggests a systematic and robust difference rather than random noise.

Together, these results reveal a consistent pattern between the presence of causal language in post titles and the structural properties of the resulting discussions. Posts with causal titles are associated with larger, deeper, and slightly broader conversation trees, suggesting that this linguistic framing co-occurs with more extensive and layered user engagement.

\section{RQ2. Temporal Dynamics of Conversation Structures}

\begin{figure}[tbhp]
\centering

\begin{subfigure}[b]{\linewidth}
    \centering
    \includegraphics[width=0.95\linewidth]{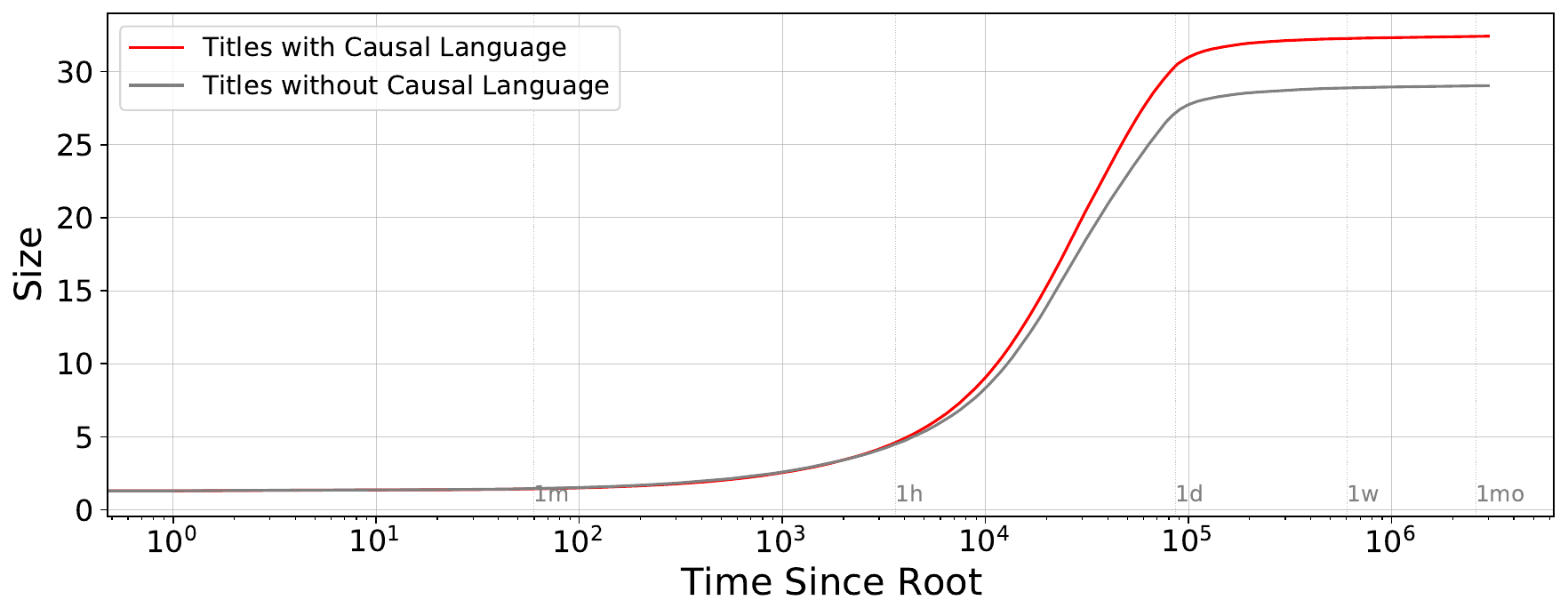}
    \caption{Cumulative growth in \textbf{size}, divergence begins around \textbf{3,000} seconds (50 minutes) after posting, indicating faster early comment accumulation for causal titles.}
    \label{fig:RQ3_time_size}
\end{subfigure}

\vspace{1em}

\begin{subfigure}[b]{\linewidth}
    \centering
    \includegraphics[width=0.95\linewidth]{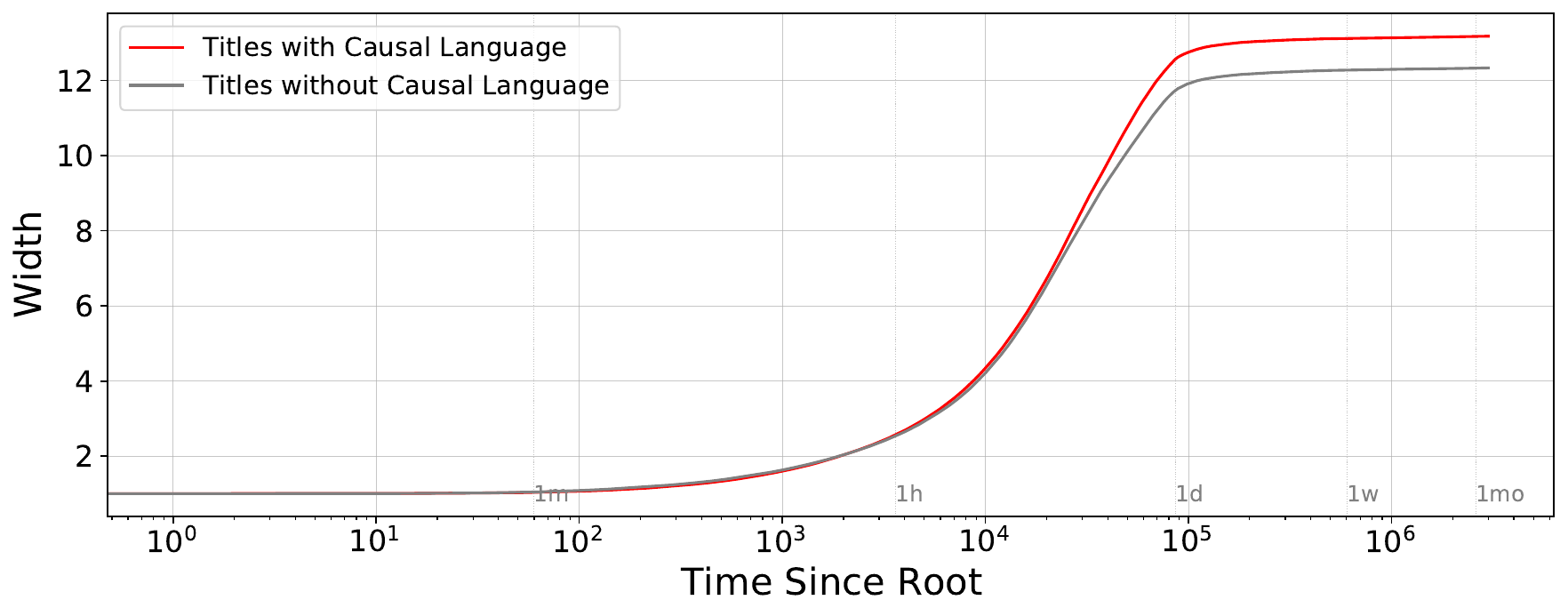}
    \caption{Cumulative growth in \textbf{width}, divergence becomes apparent around \textbf{20,000} seconds (5.55 hours), reflecting more modest and delayed differences in width engagement.}
    \label{fig:RQ3_time_width}
\end{subfigure}

\vspace{1em}

\begin{subfigure}[b]{\linewidth}
    \centering
    \includegraphics[width=0.95\linewidth]{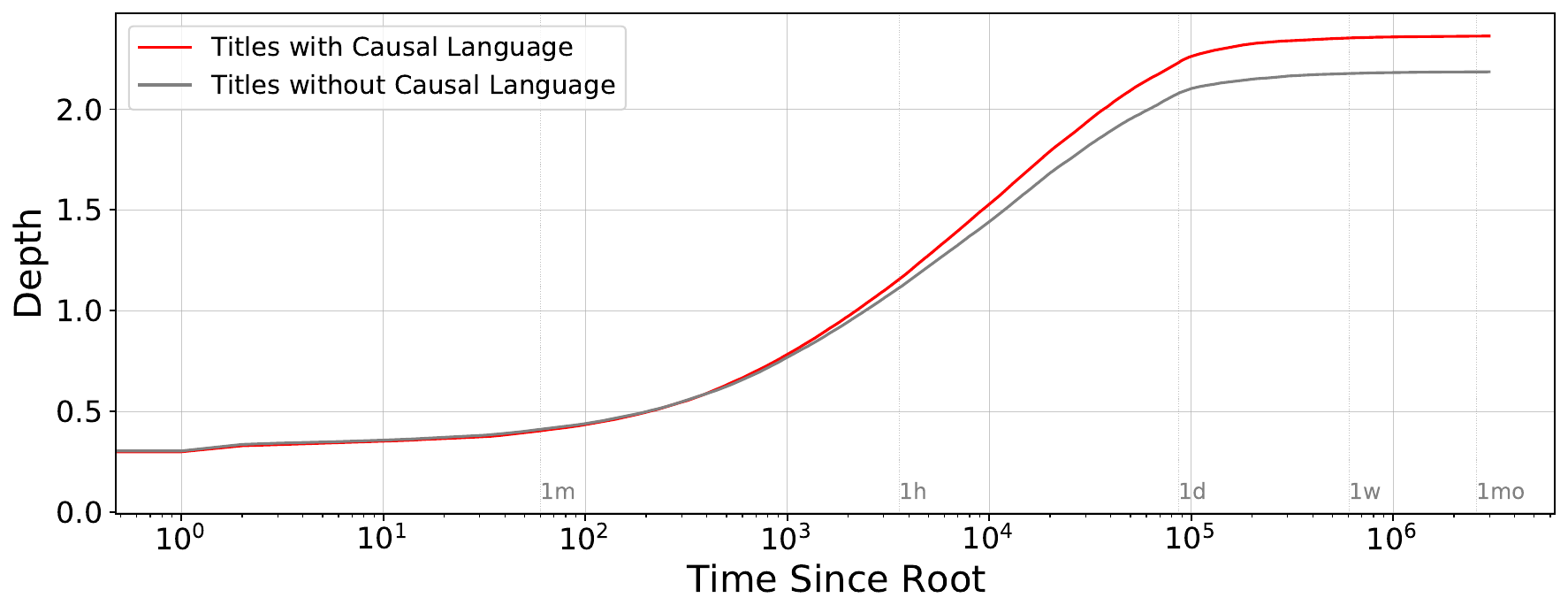}
    \caption{Cumulative growth in \textbf{depth}, divergence occurs very early, around \textbf{400} seconds, suggesting that causal framing rapidly triggers deeper interaction structures.}
    \label{fig:RQ3_time_depth}
\end{subfigure}

\caption{Temporal evolution of structural properties of discussion trees for posts with and without causal language in titles. Depth diverges first, followed by size, and then width.}
\label{fig:RQ3_time}
\end{figure}

Building on our earlier findings, we now examine how the structural properties of discussion trees evolve over time, comparing posts with and without causal language in their titles. Specifically, we track the cumulative growth of three key metrics (average size, width, and depth) as a function of time since the post was made, which lasts for  3,000,000 seconds, slightly more than a month (30 days = 2,592,000 seconds). These dynamics are visualized in Figure~\ref{fig:RQ3_time}, which plots the evolution of each structural property over time, across 57,600 posts we obtain after counterfactual control from RQ2.

Posts with causal titles accumulate comments more rapidly than non-causal posts, particularly during the early engagement window (approximately 3,000 seconds after posting). The Kolmogorov-Smirnov (K-S) test indicates a significant difference between the two trends ($D = 0.1867$, $p < 1.74 \times 10^{-47}$). This initial divergence suggests that causal framing may act as an early attention signal, prompting faster uptake and community response. Notably, this advantage persists over time: posts with causal language ultimately achieve larger final tree sizes, consistent with our earlier  analysis.

The growth in width, measured as the maximum number of sibling comments at any level, also diverges over time between the two groups, approximately 20,000 seconds after posting. The K-S test showes strong evidence of a difference between the two trends
($D = 0.1631$, $p < 2.79 \times 10^{-36}$). While both causal and non-causal threads expand in breadth during the early stages, those with causal framing show a slightly faster rise in width before plateauing. This indicates that causal language may modestly increase parallel engagement early in the discussion, prompting more users to respond directly to the post or to early comments.

The depth difference is also pronounced. As shown in Figure~\ref{fig:RQ3_time_depth}, discussion trees associated with causal language consistently develop deeper structures over time. The K-S test reveals a significant difference between the two trends ($D = 0.1835$, $p < 7.36 \times 10^{-46}$). The divergence becomes especially clear after 400 seconds, where the cumulative depth of causal threads continues to grow at a higher rate. This sustained growth suggests that causal framing not only has more engagement but also has more continued, layered interactions, supporting the interpretation that posts with causal language spark deeper conversational trajectories.

Taken together, these temporal patterns indicate that causal framing influences not only the final structure of conversations but also the trajectory by which they unfold. Notably, the divergence in structural properties emerges at different times: depth shows the earliest divergence, which begins around 400 seconds, suggesting that posts with causal framing have faster, deeper, and more layered exchanges. Size diverges next, with a clear gap appearing at 3,000 seconds, reflecting faster overall accumulation of engagement. Width diverges later, around 20,000 seconds, pointing to more modest effects on parallel participation. This staggered emergence of differences implies that causal language primarily drives early depth and sustained engagement, which subsequently supports broader participation. Overall, these findings suggest that causal framing serves as a critical initial condition shaping not just how much people engage, but how they engage, leading to deeper, more recursive conversational dynamics over time.

\section{RQ3. Early Emotion of Comment Trees}

One possible explanation for the structural differences observed in earlier sections is that causal language elicits stronger emotional responses, which in turn may drive deeper and more sustained discussions. Prior work has consistently shown that emotional expression increases attention, participation, and information diffusion online. For example, emotionally charged content is more likely to be engaged with and shared \cite{berger2012makes}, and moral-emotional language has been shown to accelerate diffusion through affective contagion and social reinforcement mechanisms \cite{brady2017emotion}.

Beyond that, research on conversation dynamics suggests that early signals play a disproportionate role in determining long-term structural outcomes. Early engagement can trigger cumulative advantage processes, where initial reactions shape visibility and participation trajectories consistent with Matthew Effects \cite{muchnik2013social, cheng2014can}. Emotional expressions in early comments may further amplify these dynamics by establishing affective norms, inviting agreement or conflict, and motivating reply behavior through arousal or controversy. Prior studies of online discussions and social media have found that emotional language, especially high-arousal emotions such as anger or surprise, can propagate through reply chains and contribute to longer and more active conversations via emotional contagion and feedback loops \cite{stieglitz2013emotions, brady2017emotion}.

Taken together, these works suggest a plausible pathway by which causal language might indirectly produce deeper and more sustained discussion structures by first eliciting stronger emotional reactions in early replies. To investigate this possibility, we analyze the emotional content of early comments in posts with and without causal language.

\begin{figure}[tbhp]
\centering
\includegraphics[width=.95\linewidth]{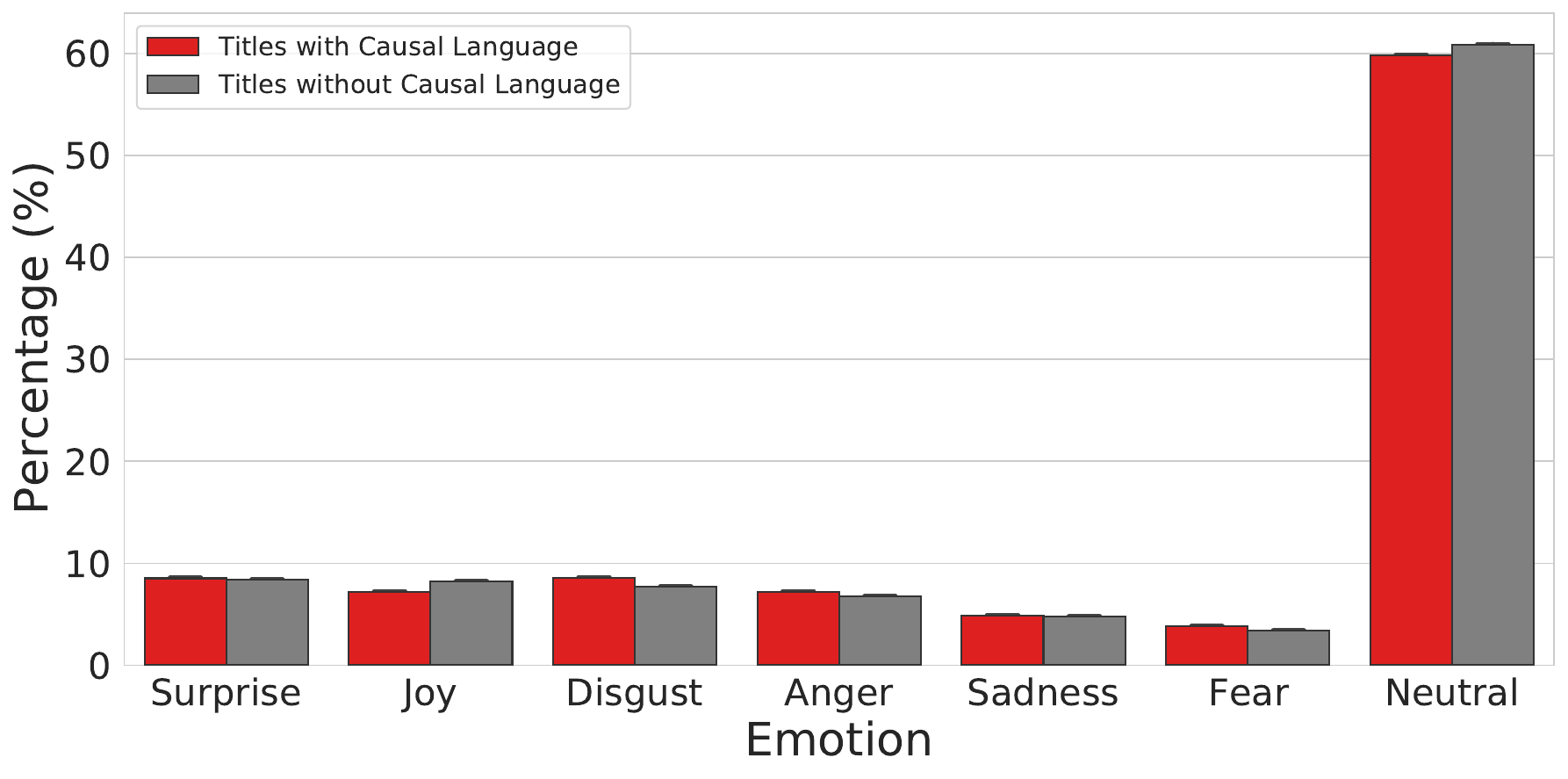}
\caption{Comparative analysis of the percentage distribution of dominant emotion in the first 10 comments in temporal order in each post, representing the earliest user reactions to the post, segregated by the presence or absence of causal language. Error bars reflect the variability across 100 bootstrapped runs. The emotional profiles of early comments are remarkably similar across posts with and without causal framing.  }
\label{fig:RQ4_emo_top10}
\end{figure}

Specifically, for each post in our dataset obtained after counterfactual experiment from RQ1, we extract the first 10 comments in temporal order, which represents the earliest user reactions to the post. We apply a pre-trained emotion classification model \cite{hartmann2022emotionenglish} to label each comment into one of seven emotion categories: anger, disgust, fear, joy, sadness, surprise, and neutral. Figure~\ref{fig:RQ4_emo_top10} displays the proportion of comments in each emotion category for both groups.

Overall, the emotional profiles of early comments are remarkably similar across posts with and without causal framing. The Mann–Whitney U test suggests no significant difference between the two groups, indicating that their distributions are highly similar (U = 26.0, p = 0.9015). In both groups, neutral comments dominate the early discourse, accounting for approximately \textit{60\%} of all replies. The relative proportions of emotional categories such as anger, joy, and sadness remain nearly identical. For example, the proportion of surprise-related comments is \textit{8.40\%} for causal posts and \textit{8.43\%} for non-causal posts; joy is \textit{7.21\%} vs. \textit{7.99\%}, respectively.

These findings suggest that early emotional intensity is not a distinguishing factor in the development of deeper or more complex conversation structures when comparing post titles with or without causal language. The similarity in emotional profiles implies that it is not heightened emotion that leads to the deeper discussions observed in causal posts. Instead, the structural divergence appears to stem more directly from the framing properties of causal language itself, and independent of emotional provocation.

\section{RQ4. Emotion of Entire Comment Trees}

While RQ3 focused on the emotional tone of early replies, here we investigate whether the overall emotion of entire discussions differs between posts with and without causal language in their titles. This analysis allows us to assess whether causal framing is associated with more emotionally charged conversations at scale, rather than just in the early moments of engagement.

For each post from the dataset we obtained after counterfactual experiment from RQ1, we aggregate the emotional classifications of all associated comments. We use the same language model-based emotion classifier \cite{hartmann2022emotionenglish} as used in RQ3 to label each comment into one of seven emotion categories: anger, disgust, fear, joy, sadness, surprise, and neutral.

\begin{figure}[tbhp]
\centering
\includegraphics[width=.95\linewidth]{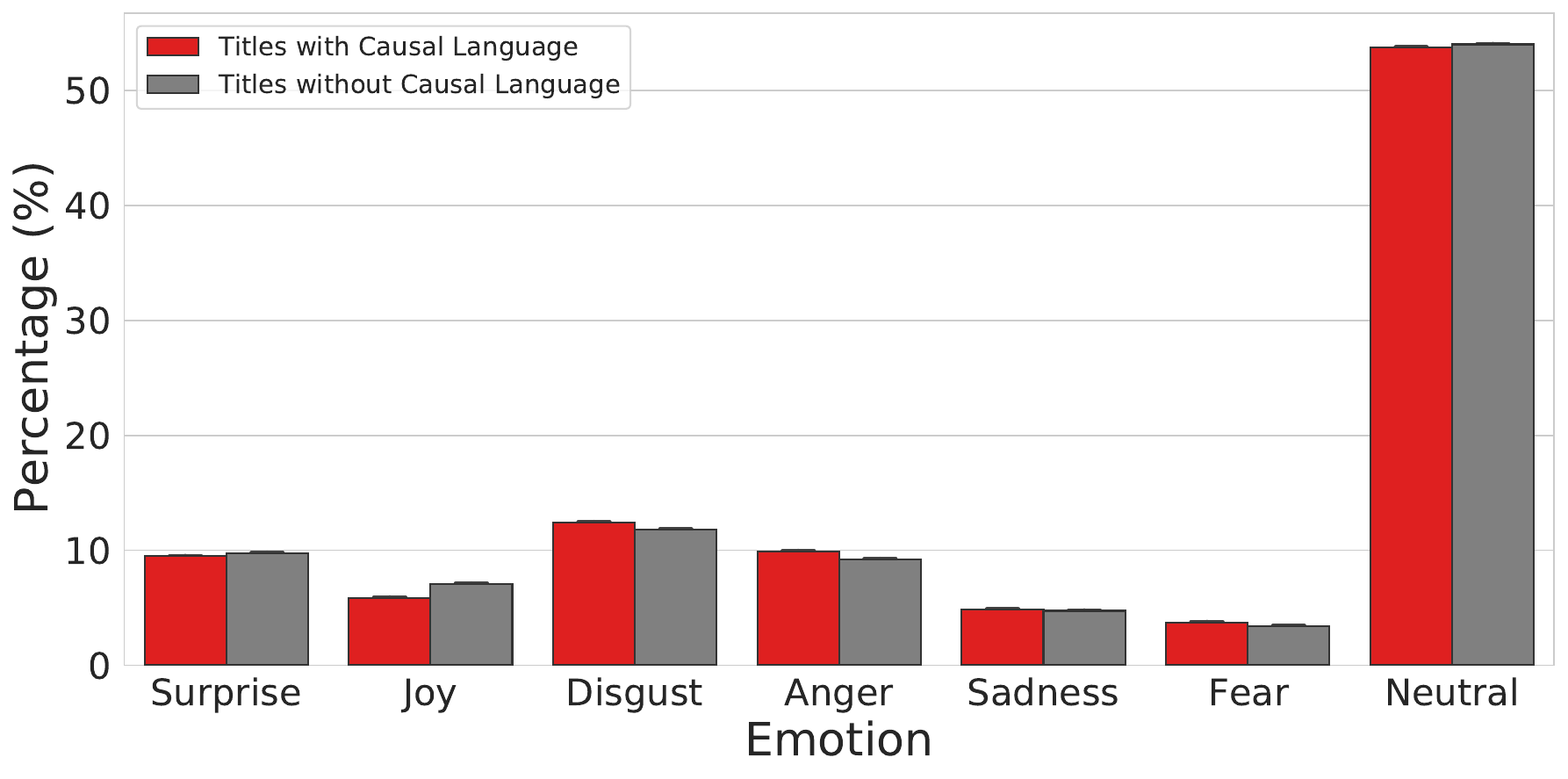}
\caption{Comparative analysis of the percentage distribution of dominant emotion in all comments, segregated by the presence or absence of causal language. Error bars reflect the variability across 100 bootstrapped runs. The emotion of comments are similar across posts with and without causal framing.}
\label{fig:RQ5_emo}
\end{figure}

Figure~\ref{fig:RQ5_emo} presents the distribution of emotion categories across the full set of comments for causal and non-causal posts.  Consistent with our findings in RQ3, we observe no substantial divergence in the overall emotional composition of discussions. The Mann–Whitney U test indicates that the two groups are very similar (U = 27.0, p = 0.8048). Both groups are dominated by neutral comments, accounting for approximately \textit{54\%} of all replies. The relative prevalence of emotional categories remains similar across groups. For example, the proportion of surprise-related comments is \textit{9.47\%} for causal posts and \textit{9.78\%} for non-causal posts; joy is \textit{5.85\%} vs. \textit{7.08\%}; and disgust is \textit{12.40\%} vs. \textit{11.76\%}.

These results suggest that even across the full temporal span of the conversation, causal framing does not systematically trigger more emotionally expressive discussions. This suggests the interpretation that the depth and complexity of conversations in causal threads arise from causal framing itself rather than from emotion-driven escalation.

\section{RQ5. Qualitative Analysis}

To further explore potential mechanisms underlying the observed structural differences beyond emotional amplification, we conduct a lightweight qualitative analysis of early reply-to-reply interactions. While prior sections test and rule out affective pathways as a primary driver of deeper conversational structures, other alternative mechanisms remain plausible. Prior work suggests that sustained conversational depth can emerge not only from affective escalation but also from cognitive and interactional dynamics such as explanation-seeking, interpretation, and experience sharing \cite{lombrozo2006structure, graesser1994constructing, zhang2018characterizing}. In particular, causal framing may activate a cognitive orientation toward understanding “why” and “how,” prompting users to articulate mechanisms, propose hypotheses, or refine causal interpretations. Such contributions often require elaboration and invite follow-up responses, naturally giving rise to multi-turn exchanges.

In addition, causal titles may encourage users to draw on personal experiences as informal evidence, embedding individual narratives within broader explanatory threads. This process can support iterative sensemaking, where participants compare, validate, or challenge each other’s accounts. Previous research highlights that disagreement, requests for clarification, and elaboration, rather than emotional intensity alone, are key drivers of sustained interaction and conversational depth in online discussions \cite{danescu2013computational, kumar2010dynamics}. These interactional moves create dependencies between turns, increasing the likelihood that discussions evolve into deeper, branching structures.

We conduct a lightweight qualitative analysis of reply-to-reply interactions sampled from causal and non-causal posts after counterfactual matching, allowing each response to be interpreted in its local conversational context.  We randomly selected 30 matched pairs of causal and non-causal post titles after counterfactual experiment and extracted their associated comment trees. To ensure that the sampled threads contained sufficient conversational depth for meaningful interactional analysis, we restricted sampling to pairs where both trees have at least two reply-to-reply comment threads (i.e., at least two leafs at depth $\geq$ 2). From each tree, we selected the first two reply-to-reply comment threads in temporal order and took the first two comments from each thread. This yielded a set of comments that captures the early deepening of discussion, with 191 comments in total, of which 95 are from causal trees and 96 are from non-causal trees.

Each reply-to-reply comment was then coded according to its primary conversational function into one of three categories: explanation or reasoning, experience or anecdote, and others. This design allows us to examine whether causal framing shapes how users reason, mobilize personal experience, and negotiate claims within early deep interactional sequences, rather than merely affecting emotional tone. This scheme directly probes mechanisms suggested in prior work: explanation-oriented reasoning reflects causal sensemaking \cite{lombrozo2006structure}, and experience sharing serves as informal evidence in collective reasoning \cite{pennington1992explaining, zhang2018characterizing}. This analysis provides convergent qualitative evidence that causal framing may foster deeper conversational structures by promoting explanation-seeking and iterative reasoning, rather than by amplifying emotional responses.

Our qualitative analysis reveals systematic differences in the types of interactions that emerge in deeper portions of causal versus non-causal discussion trees. In posts with causal framing, reply-to-reply comments more frequently involved reasoning and explanation (44.21\% in causal posts, compared to  19.79\% in non-causal posts), including attempts to articulate mechanisms, propose causal pathways, or refine prior explanations (e.g., ``Talib wasn't mad at Crabtree specifically, he was mad because his position coach gave him shit after the Raiders beat the Broncos.”). These comments often built directly on their parent comments by elaborating on proposed causes, introducing additional factors, or clarifying how one factor might lead to another.

Experience sharing also differed systematically between causal and non-causal conditions. In posts with causal framing, personal experiences appeared more frequently in comments and were more tightly integrated into ongoing explanatory threads  (13.68\%). Contributors often presented their own experiences as evidence to support, refine, or contest proposed causal claims (e.g., ``Taking a cab in Cairo is a life altering experience: As an atheist I have never prayed harder than when using a cab here. I'm still here so someone or thing heard my prayer. :-)”), which in turn elicited follow-up questions, counterexamples, or additional reasoning. In matched non-causal posts, experiential comments were less frequent (6.25\%). This pattern suggests that causal framing encourages users to mobilize personal experience as part of collective sensemaking, thereby promoting deeper conversational trajectories rather than merely parallel contributions.

\section{Discussion}

Our study investigates how the presence of causal language in post titles shapes the structure and dynamics of online conversations. Across five research questions, we find consistent evidence that causal framing correlates with deeper, more participatory, and more temporally sustained discussions. In this section, we show the theoretical implications of these findings, discuss their implications to the larger CSCW community, and reflect on methodological limitations and opportunities for future work.

\subsection{Causal Framing as a Structural Signal}

Our results suggest that causal language functions not merely as a stylistic or rhetorical device, but as a powerful conversational signal that shapes the architecture of online discourse. In RQ1, we show that posts with causal language are associated with significantly larger and deeper comment trees, showing higher engagement and layered interaction, after controlling for potential confounds through counterfactual matching. These findings suggest that the presence of causal framing is strongly associated with the way conversations become at the structural levels.

Causal language can drive extended interaction. Unlike emotionally provocative language or clickbait phrasing, which often lead to bursts of shallow attention, causal framing may invite deeper participation, where users engage not only with the content but with the underlying reasoning implied by the title. This  suggests that how a question is asked, or a topic is framed, can profoundly shape how communities respond.

\subsection{Temporal Dynamics and Early Shaping Effects}

Our temporal analysis (RQ2) further supports the idea that causal framing exerts influence early in the lifecycle of a conversation. Posts with causal language show faster growth in size, width, and depth, especially during the initial period after posting. This early divergence suggests that causal framing prime users not just to respond, but to engage in more extended interaction. Rather than simply amplifying overall volume, it shapes the pacing and layering of discourse from the beginning.

These temporal findings highlighting how early design elements, such as the causal language in a title, can steer long-term interactional outcomes. From a design perspective, this opens opportunities for platforms to think more intentionally about how causal framing affects discussion quality, not just quantity.

\subsection{On the Practical Significance of Effect Sizes}

A key question raised by our results is how to interpret the magnitude of the observed effects. As shown in Section 6, the Kolmogorov–Smirnov statistics comparing causal and non-causal titles are numerically small (Figure ~\ref{fig:RQ2_CF_144_200}). These values might appear modest, especially when contrasted with effect size conventions from experimental psychology. However, such conventions are poorly suited for highly skewed, heavy-tailed systems such as online discussion networks.

Reddit discussions are not normally distributed. As shown in Figure~\ref{fig:RQ2_CF_144_200}, most posts receive very little engagement, while a small fraction grow into extremely large and deeply nested trees. In such systems, the meaningful unit of change is not the average post but the probability of entering a growth trajectory that produces sustained interaction. Small distributional shifts in this regime can translate into large differences in the number of participants, reply pathways, and conversational persistence.

This can be seen directly in our percentile analyses. At the 99th percentile, posts with causal titles receive 573 comments versus 504 for non-causal titles (Figure ~\ref{fig:RQ2_CF_144_200_size}), a difference of nearly 70 additional contributions. Likewise, causal posts reach a maximum depth of 17 versus 16 (Figure~\ref{fig:RQ2_CF_144_200_depth}) and an average depth of 4.38 versus 4.14 (Figure~\ref{fig:RQ2_CF_144_200_avg_depth}). Although these differences appear numerically small, a one-unit increase in depth in a reply tree represents the survival of an additional conversational generation: more opportunities for users to reply to one another, introduce new evidence, and branch into new subthreads. In practical terms, this corresponds to dozens of additional participants and hundreds of additional potential interaction paths in large threads.

The increase in average depth indicates that a larger fraction of comment chains extend into sustained back-and-forth interaction rather than terminating early. This shift implies more opportunities for clarification, disagreement, and iterative reasoning within threads. Consistent with our qualitative analysis (Section 10), these deeper chains are often characterized by explanation-building and experience sharing, where users refine causal claims or introduce personal evidence in response to others.

At scale, these small shifts accumulate: in large threads, they translate into dozens of additional reply-to-reply interactions and substantially more branching conversational pathways. Importantly, this does not simply increase participation volume, but changes the structure of engagement, making discussions more recursive, interactive, and collectively constructed.

The temporal results further show that these differences are not confined to the tail of the distribution but arise from early divergence in growth dynamics. As Figure 4 shows, depth diverges within approximately 400 seconds, size around 3,000 seconds, and width around 20,000 seconds. This staggered pattern implies that causal framing alters whether early replies turn into recursive, multi-turn exchanges, which then compound into larger and more persistent structures. In heavy-tailed systems, such early perturbations have disproportionate long-term consequences: once a thread begins to accumulate depth, cumulative advantage mechanisms increase its visibility and attractiveness to new participants.

These effects matter because community dynamics are not driven by the median post but by the tails of the distribution. Large, persistent threads dominate what users see, what norms are reinforced, and what knowledge is produced. A framing mechanism that systematically nudges threads toward deeper and longer-lived trajectories, even by a few percent, can therefore reshape the discursive ecology of a community at scale.

In this sense, causal language functions less like a blunt engagement booster and more like a structural catalyst: it does not create activity from nothing, but it increases the probability that early interaction evolves into sustained, multi-turn exchange. That property is precisely what distinguishes shallow chatter from meaningful online discussion.

\subsection{Structural Depth, Not Emotional Escalation}

One plausible alternative explanation was that posts with causal language provoke stronger emotional responses, which then lead to deeper discussions via affective contagion or the Matthew Effect. However, our analyses in RQ3 and RQ4 do not support this hypothesis. The emotional profiles of both early and full-thread comments are strikingly similar between causal and non-causal posts, with neutral content dominating across the board. This suggests that the structural divergence we observe is not driven by affective intensity but is more directly attributable to the informational and inferential qualities embedded in causal framing itself.

Our findings also add to another layer to a dominant strand of social media research that treats emotional arousal as the primary driver of engagement. Work on moral-emotional language and virality suggests that anger, outrage, and surprise amplify diffusion and participation \cite{berger2012makes, stieglitz2013emotions, brady2017emotion}. Yet we observe that causal framing increases depth and persistence without increasing emotionality. This indicates a distinct pathway to engagement: not affective contagion, but cognitive elaboration and explanation-seeking. In doing so, our results complement and qualify emotion-centric theories of online engagement by identifying a structurally important non-emotional mechanism.

The qualitative analysis in Section 10 provides convergent evidence for this interpretation. Although causal and non-causal threads do not differ in emotional tone, their deeper conversational dynamics are systematically different. In causal threads, early reply-to-reply interactions more often involve explanation, experience-based evidence, and stance-taking in response to prior claims, whereas non-causal threads more frequently contain brief or terminal responses. These interactional patterns help explain why causal framing produces greater depth: it invites users to reason about mechanisms and supply personal evidence, which are activities that naturally generate multi-turn exchanges.

This distinction is critical. It indicates that deeper conversations can emerge not just from emotional provocation, but from cognitive framing. In doing so, our findings suggest that beyond some prevailing findings in social media research \cite{brady2017emotion, stieglitz2013emotions}, which often prioritize emotional valence as a key predictor of engagement, the role of causal language is a structurally influential linguistic feature.

\subsection{Implications}

A large body of CSCW and network science research has shown that online discussions grow through reinforcement and early-activity cascades. Generative models of threads describe how replies beget replies through preferential attachment and cumulative advantage \cite{gomez2008statistical, kumar2010dynamics, gomez2013likelihood, medvedev2019modelling}. Empirical work has similarly demonstrated that early replies, timing, and platform affordances determine whether threads collapse or become long-lived \cite{krohn2019modelling, horawalavithana2022online, aragon2017generative, budak2017threading}.

Our findings extend this literature by identifying linguistic causal framing as a previously unmeasured upstream driver of these growth dynamics. Whereas prior work treats early replies as exogenous signals, we show that the probability of recursive replying is partly shaped by how the original post frames its content in causal terms. The fact that depth diverges within 400 seconds (Figure \ref{fig:RQ3_time_depth}) suggests that causal framing alters the initial branching probability of replies, a parameter that most existing generative models leave unspecified.

In this sense, causal language functions analogously to the social catalysts identified by Saveski et al. \cite{saveski2021social}, but at the level of text rather than users. It creates conditions under which participants are more likely to reply to one another rather than only to the root post. This shifts threads from star-shaped to chain- and tree-like structures, producing more layered interaction and collective reasoning.

Our findings suggest that causal framing offers a powerful yet underutilized design lever for shaping online discourse. Subtle linguistic features, especially causal phrasing, can systematically influence how communities engage with a post, independent of its topic or emotional charge. This indicates that interaction depth is not only a function of content but also of framing, raising questions about how current platforms privilege metrics of surface-level engagement while neglecting conversational quality.

At the level of concrete interventions, platforms could experiment with lightweight nudges that encourage explanation-oriented contributions. Posting interfaces might, for instance, suggest causal formulations in titles or initial prompts, particularly in deliberation-heavy contexts such as science, news, or policy subreddits. These nudges could take the form of optional templates, context-sensitive hints, or community-specific guidelines. Care must be taken, however, to prevent superficial or manipulative uses of causal phrasing that increase apparent engagement while degrading epistemic quality or promoting unsupported causal claims.

Beyond these immediate design tweaks, platforms must also consider how such interventions interact with their underlying incentive structures. If algorithms continue to reward clicks, shares, or raw activity over meaningful exchange, the benefits of causal framing nudges may remain limited. Effective implementation therefore requires alignment between interface-level features and platform-wide governance priorities, ensuring that deeper forms of engagement are not inadvertently penalized by existing ranking or moderation systems. In this sense, causal framing could serve as a diagnostic tool as much as an intervention. It can be a way to measure whether platforms are genuinely fostering deliberation rather than merely amplifying attention.

At a broader level, these findings invite rethinking moderation not only as a reactive process of filtering harmful content but as a proactive shaping of conversational norms. By making visible the subtle ways language guides interaction, platforms can expand the scope of moderation from preventing harm to cultivating more reflective, knowledge-rich dialogue. Such a shift would align with emerging perspectives in human-computer interaction and computer-supported cooperative work that emphasize design for long-term relational and epistemic outcomes, rather than short-term engagement alone.

Finally, the broader societal implications of these findings extend beyond platform design. Causal framing not only alters the immediate tone of a conversation but also shapes its trajectory, determining whether discussions dissipate quickly or mature into sustained exchanges. By recognizing language as a structural force rather than a neutral medium, platforms can foster healthier conditions for knowledge exchange, collaborative reasoning, and democratic deliberation in networked publics. For the research community, this highlights the value of studying framing as a key intersection between cognition, interaction, and sociotechnical design.

\subsection{Limitations and Ethical Considerations}

While our findings offer robust insights into how causal framing relates to online discourse, several limitations must be acknowledged.

First, our analysis is based exclusively on Reddit, which has its own cultural norms, interface design, and community moderation practices. Reddit’s threaded discussion format and subreddit-based structure encourage public, often asynchronous conversations, which may differ substantially from dynamics on platforms like Twitter, Facebook, or TikTok, where interactions are often shorter, less nested, or algorithmically mediated in different ways. This raises questions about generalizability: linguistic framing effects observed here may manifest differently in platforms with distinct social, technical, or affective architectures.

Second, our sample is restricted to the top 200 subreddits by engagement in 2023, which biases the dataset toward large, active, and relatively mainstream communities and timing. Niche or marginalized spaces may have different norms around language use and engagement that are not captured in our analysis.

Third, while we rely on automated methods for emotion detection and semantic similarity, these models are not perfect. Emotion classifiers may miss subtle affective cues such as sarcasm, irony, or culturally specific expressions. 

From an ethical perspective, while Reddit is a public platform, user-generated content still warrants careful handling. We avoid any analysis of identifiable user behavior and report only aggregated patterns at the community level to protect privacy and consent.

\subsection{Future Work}

Our findings open up several promising directions for future research. One key avenue is to extend this analysis across platforms with different conversational structures. For example, platforms like YouTube or Facebook offer interactions that may respond differently to causal framing. 

Another direction involves examining more granular linguistic features beyond causal framing. Future work could explore how combinations of stance, uncertainty, or question type interact with causal language to shape user response. Additionally, incorporating user-level variables such as reputation, or prior activity could reveal how audience characteristics mediate framing effects.

Finally, future research could integrate experimental or quasi-experimental methods, such as A/B testing post titles or analyzing natural experiments triggered by title framing, to more directly assess the mechanisms through which linguistic framing affects discourse. Longitudinal analyses could also explore whether the use of causal framing has changed over time, and how such shifts relate to broader trends in public discourse.

Together, these future directions can help deepen our understanding of the subtle yet powerful ways that causal language shapes not only what and how we say online, but how others respond.

\section{Conclusion}

This study examines how causal language in post titles shapes the structure, trajectory, and emotional tone of online discussions. Using large-scale Reddit data, we show that posts with causal language consistently generate larger, deeper, and more sustained conversation trees compared to non-causal counterparts. These patterns persist even after counterfactual experiment, highlighting a robust relationship between linguistic framing and conversational structure. Moreover, we find that the structural divergence begins early: causal posts accumulate depth and size more quickly, suggesting that framing influences not only eventual outcomes but also the temporal trajectory of engagement.

Importantly, we find that this structural divergence is not driven by emotional amplification: the emotional profiles of both early and full-thread comments remain largely similar between causal and non-causal posts. Instead, our findings suggest that causal language acts as a discourse-level signal that invites layered, deliberative engagement. The qualitative analysis shows that causal framing posts more often invite reasoning and explanatory comments, as well as experience-sharing comments, in early reply-to-reply interactions, which helps explain why causal framing produces greater depth.

By combining counterfactual design, structural analysis, temporal dynamics, emotion modeling, and qualitative analysis, our work advances understanding of how subtle linguistic choices shape the collective behaviors of online communities. These insights contribute to broader research community for considerations about platform design, user engagement, and the sociotechnical forces that govern how people interact, reason, and respond in digital spaces.

\begin{acks}
This research was supported, in part, by AFOSR MURI grant \#FA9550-22-1-0380. 
\end{acks}

\bibliographystyle{ACM-Reference-Format}
\bibliography{main}

@String{Computing = "Computing" }

@String{Academic = "Academic Press" }

@String{Springer = "Springer-Verlag" }

@book{mitchell2009complexity,
  title={Complexity: A guided tour},
  author={Mitchell, Melanie},
  year={2009},
  publisher={Oxford university press}
}

@book{strogatz2024nonlinear,
  title={Nonlinear dynamics and chaos: with applications to physics, biology, chemistry, and engineering},
  author={Strogatz, Steven H},
  year={2024},
  publisher={Chapman and Hall/CRC}
}

@inproceedings{priniski2023pipeline,
  title={Pipeline for modeling causal beliefs from natural language},
  author={Priniski, John and Verma, Ishaan and Morstatter, Fred},
  booktitle={Association for Computational Linguistics (Volume 3: System Demonstrations)},
  pages={436--443},
  year={2023}
}

@article{liu2019roberta,
  title={Roberta: A robustly optimized bert pretraining approach},
  author={Liu, Yinhan and Ott, Myle and Goyal, Naman and Du, Jingfei and Joshi, Mandar and Chen, Danqi and Levy, Omer and Lewis, Mike and Zettlemoyer, Luke and Stoyanov, Veselin},
  journal={arXiv preprint arXiv:1907.11692},
  year={2019}
}

@article{stuart2010matching,
  title={Matching methods for causal inference: A review and a look forward},
  author={Stuart, Elizabeth A},
  journal={Statistical science: a review journal of the Institute of Mathematical Statistics},
  volume={25},
  number={1},
  pages={1},
  year={2010},
  publisher={NIH Public Access}
}

@inproceedings{reimers-2019-sentence-bert,
  title={Sentence-BERT: Sentence Embeddings using Siamese BERT-Networks},
  author={Reimers, Nils and Gurevych, Iryna},
  booktitle={Proceedings of the 2019 Conference on Empirical Methods in Natural Language Processing and the 9th International Joint Conference on Natural Language Processing (EMNLP-IJCNLP)},
  pages={3982--3992},
  year={2019}
}

@article{muchnik2013social,
  title={Social influence bias: A randomized experiment},
  author={Muchnik, Lev and Aral, Sinan and Taylor, Sean J},
  journal={Science},
  volume={341},
  number={6146},
  pages={647--651},
  year={2013},
  publisher={American Association for the Advancement of Science}
}

@article{ekman1999basic,
  title={Basic emotions},
  author={Ekman, Paul},
  journal={Handbook of cognition and emotion},
  volume={98},
  number={45-60},
  pages={16},
  year={1999}
}

@misc{hartmann2022emotionenglish,
  author={Hartmann, Jochen},
  title={Emotion English DistilRoBERTa-base},
  year={2022},
  howpublished = {\url{https://huggingface.co/j-hartmann/emotion-english-distilroberta-base/}},
}

@inproceedings{cheng2014can,
  title={Can cascades be predicted?},
  author={Cheng, Justin and Adamic, Lada and Dow, P Alex and Kleinberg, Jon Michael and Leskovec, Jure},
  booktitle={Proceedings of the 23rd international conference on World wide web},
  pages={925--936},
  year={2014}
}

@article{brady2017emotion,
  title={Emotion shapes the diffusion of moralized content in social networks},
  author={Brady, William J and Wills, Julian A and Jost, John T and Tucker, Joshua A and Van Bavel, Jay J},
  journal={Proceedings of the National Academy of Sciences},
  volume={114},
  number={28},
  pages={7313--7318},
  year={2017},
  publisher={National Academy of Sciences}
}

@article{berger2012makes,
  title={What makes online content viral?},
  author={Berger, Jonah and Milkman, Katherine L},
  journal={Journal of marketing research},
  volume={49},
  number={2},
  pages={192--205},
  year={2012},
  publisher={SAGE Publications Sage CA: Los Angeles, CA}
}

@article{redditdata,
title= {Reddit comments/submissions 2005-06 to 2023-12},
journal= {},
author= {{stuck\_in\_the\_matrix, Watchful1, RaiderBDev}},
year= {},
url= {},
terms= {},
license= {},
superseded= {https://academictorrents.com/details/ba051999301b109eab37d16f027b3f49ade2de13}
}

@inproceedings{gomez2008statistical,
  title={Statistical analysis of the social network and discussion threads in slashdot},
  author={G{\'o}mez, Vicen{\c{c}} and Kaltenbrunner, Andreas and L{\'o}pez, Vicente},
  booktitle={Proceedings of the 17th international conference on World Wide Web},
  pages={645--654},
  year={2008}
}

@article{danescu2013computational,
  title={A computational approach to politeness with application to social factors},
  author={Danescu-Niculescu-Mizil, Cristian and Sudhof, Moritz and Jurafsky, Dan and Leskovec, Jure and Potts, Christopher},
  journal={arXiv preprint arXiv:1306.6078},
  year={2013}
}

@inproceedings{kumar2010dynamics,
  title={Dynamics of conversations},
  author={Kumar, Ravi and Mahdian, Mohammad and McGlohon, Mary},
  booktitle={Proceedings of the 16th ACM SIGKDD international conference on Knowledge discovery and data mining},
  pages={553--562},
  year={2010}
}

@article{gomez2013likelihood,
  title={A likelihood-based framework for the analysis of discussion threads},
  author={G{\'o}mez, Vicen{\c{c}} and Kappen, Hilbert J and Litvak, Nelly and Kaltenbrunner, Andreas},
  journal={World Wide Web},
  volume={16},
  pages={645--675},
  year={2013},
  publisher={Springer}
}

@article{medvedev2019modelling,
  title={Modelling structure and predicting dynamics of discussion threads in online boards},
  author={Medvedev, Alexey N and Delvenne, Jean-Charles and Lambiotte, Renaud},
  journal={Journal of Complex Networks},
  volume={7},
  number={1},
  pages={67--82},
  year={2019},
  publisher={Oxford University Press}
}

@article{medvedev2017anatomy,
  title={The anatomy of Reddit: An overview of academic research},
  author={Medvedev, Alexey N and Lambiotte, Renaud and Delvenne, Jean-Charles},
  journal={Dynamics on and of Complex Networks},
  pages={183--204},
  year={2017},
  publisher={Springer}
}

@article{aragon2017generative,
  title={Generative models of online discussion threads: state of the art and research challenges},
  author={Arag{\'o}n, Pablo and G{\'o}mez, Vicen{\c{c}} and Garc{\'\i}a, David and Kaltenbrunner, Andreas},
  journal={Journal of Internet Services and Applications},
  volume={8},
  pages={1--17},
  year={2017},
  publisher={Springer}
}

@inproceedings{aragon2017thread,
  title={To thread or not to thread: The impact of conversation threading on online discussion},
  author={Arag{\'o}n, Pablo and G{\'o}mez, Vicen{\c{c}} and Kaltenbrunner, Andreaks},
  booktitle={Proceedings of the International AAAI Conference on Web and social media},
  volume={11},
  number={1},
  pages={12--21},
  year={2017}
}

@article{goglia2024structure,
  title={Structure and dynamics of growing networks of Reddit threads},
  author={Goglia, Diletta and Vega, Davide},
  journal={Applied Network Science},
  volume={9},
  number={1},
  pages={48},
  year={2024},
  publisher={Springer}
}

@article{yu2024characterizing,
  title={Characterizing the Structure of Online Conversations Across Reddit},
  author={Yu, Yulin and Jiang, Julie and Dhillon, Paramveer S},
  journal={Proceedings of the ACM on Human-Computer Interaction},
  volume={8},
  number={CSCW2},
  pages={1--23},
  year={2024},
  publisher={ACM New York, NY, USA}
}

@inproceedings{krohn2019modelling,
  title={Modelling online comment threads from their start},
  author={Krohn, Rachel and Weninger, Tim},
  booktitle={2019 IEEE International Conference on Big Data (Big Data)},
  pages={820--829},
  year={2019},
  organization={IEEE}
}

@article{horawalavithana2022online,
  title={Online discussion threads as conversation pools: predicting the growth of discussion threads on reddit},
  author={Horawalavithana, Sameera and Choudhury, Nazim and Skvoretz, John and Iamnitchi, Adriana},
  journal={Computational and Mathematical Organization Theory},
  pages={1--29},
  year={2022},
  publisher={Springer}
}

@article{gopnik2004theory,
  title={A theory of causal learning in children: causal maps and Bayes nets.},
  author={Gopnik, Alison and Glymour, Clark and Sobel, David M and Schulz, Laura E and Kushnir, Tamar and Danks, David},
  journal={Psychological review},
  volume={111},
  number={1},
  pages={3},
  year={2004},
  publisher={American Psychological Association}
}

@book{gopnik2007causal,
  title={Causal learning: Psychology, philosophy, and computation},
  author={Gopnik, Alison and Schulz, Laura},
  year={2007},
  publisher={Oxford University Press}
}

@article{keil2006explanation,
  title={Explanation and understanding},
  author={Keil, Frank C},
  journal={Annu. Rev. Psychol.},
  volume={57},
  number={1},
  pages={227--254},
  year={2006},
  publisher={Annual Reviews}
}

@book{sloman2009causal,
  title={Causal models: How people think about the world and its alternatives},
  author={Sloman, Steven and Sloman, Steven A},
  year={2009},
  publisher={Oxford University Press}
}

@article{lombrozo2006structure,
  title={The structure and function of explanations},
  author={Lombrozo, Tania},
  journal={Trends in cognitive sciences},
  volume={10},
  number={10},
  pages={464--470},
  year={2006},
  publisher={Elsevier}
}

@article{gopnik2012reconstructing,
  title={Reconstructing constructivism: causal models, Bayesian learning mechanisms, and the theory theory.},
  author={Gopnik, Alison and Wellman, Henry M},
  journal={Psychological bulletin},
  volume={138},
  number={6},
  pages={1085},
  year={2012},
  publisher={American Psychological Association}
}

@article{graesser1994constructing,
  title={Constructing inferences during narrative text comprehension.},
  author={Graesser, Arthur C and Singer, Murray and Trabasso, Tom},
  journal={Psychological review},
  volume={101},
  number={3},
  pages={371},
  year={1994},
  publisher={American Psychological Association}
}

@article{sanders1992toward,
  title={Toward a taxonomy of coherence relations},
  author={Sanders, Ted JM and Spooren, Wilbert PM and Noordman, Leo GM},
  journal={Discourse processes},
  volume={15},
  number={1},
  pages={1--35},
  year={1992},
  publisher={Taylor \& Francis}
}

@article{weiner1985attributional,
  title={An attributional theory of achievement motivation and emotion.},
  author={Weiner, Bernard},
  journal={Psychological review},
  volume={92},
  number={4},
  pages={548},
  year={1985},
  publisher={American Psychological Association}
}

@article{iyengar1987television,
  title={Television news and citizens' explanations of national affairs},
  author={Iyengar, Shanto},
  journal={American Political Science Review},
  volume={81},
  number={3},
  pages={815--831},
  year={1987},
  publisher={Cambridge University Press}
}

@article{lagnado2008judgments,
  title={Judgments of cause and blame: The effects of intentionality and foreseeability},
  author={Lagnado, David A and Channon, Shelley},
  journal={Cognition},
  volume={108},
  number={3},
  pages={754--770},
  year={2008},
  publisher={Elsevier}
}

@article{pennington1992explaining,
  title={Explaining the evidence: Tests of the Story Model for juror decision making.},
  author={Pennington, Nancy and Hastie, Reid},
  journal={Journal of personality and social psychology},
  volume={62},
  number={2},
  pages={189},
  year={1992},
  publisher={American Psychological Association}
}

@article{langer1978mindlessness,
  title={The mindlessness of ostensibly thoughtful action: The role of" placebic" information in interpersonal interaction.},
  author={Langer, Ellen J and Blank, Arthur and Chanowitz, Benzion},
  journal={Journal of personality and social psychology},
  volume={36},
  number={6},
  pages={635},
  year={1978},
  publisher={American Psychological Association}
}

@article{budak2017threading,
  title={Threading is sticky: How threaded conversations promote comment system user retention},
  author={Budak, Ceren and Garrett, R Kelly and Resnick, Paul and Kamin, Julia},
  journal={Proceedings of the ACM on Human-Computer Interaction},
  volume={1},
  number={CSCW},
  pages={1--20},
  year={2017},
  publisher={ACM New York, NY, USA}
}

@article{chandrasekharan2017you,
  title={You can't stay here: The efficacy of reddit's 2015 ban examined through hate speech},
  author={Chandrasekharan, Eshwar and Pavalanathan, Umashanthi and Srinivasan, Anirudh and Glynn, Adam and Eisenstein, Jacob and Gilbert, Eric},
  journal={Proceedings of the ACM on human-computer interaction},
  volume={1},
  number={CSCW},
  pages={1--22},
  year={2017},
  publisher={ACM New York, NY, USA}
}

@article{saveski2021social,
  title={Social catalysts: Characterizing people who spark conversations among others},
  author={Saveski, Martin and Kooti, Farshad and Morelli Vitousek, Sylvia and Diuk, Carlos and Bartlett, Bryce and Adamic, Lada A},
  journal={Proceedings of the ACM on Human-Computer Interaction},
  volume={5},
  number={CSCW2},
  pages={1--20},
  year={2021},
  publisher={ACM New York, NY, USA}
}

@article{zhang2018characterizing,
  title={Characterizing online public discussions through patterns of participant interactions},
  author={Zhang, Justine and Danescu-Niculescu-Mizil, Cristian and Sauper, Christina and Taylor, Sean J},
  journal={Proceedings of the ACM on Human-Computer Interaction},
  volume={2},
  number={CSCW},
  pages={1--27},
  year={2018},
  publisher={ACM New York, NY, USA}
}

@inproceedings{shi2024diffusion,
  title={The Diffusion of Causal Language in Social Networks},
  author={Shi, Zhuoyu and Morstatter, Fred},
  booktitle={Proceedings of the International AAAI Conference on Web and Social Media},
  volume={18},
  pages={1422--1435},
  year={2024}
}

@article{stieglitz2013emotions,
  title={Emotions and information diffusion in social media—sentiment of microblogs and sharing behavior},
  author={Stieglitz, Stefan and Dang-Xuan, Linh},
  journal={Journal of management information systems},
  volume={29},
  number={4},
  pages={217--248},
  year={2013},
  publisher={Taylor \& Francis}
}

@inproceedings{krohn2022subreddit,
  title={Subreddit links drive community creation and user engagement on reddit},
  author={Krohn, Rachel and Weninger, Tim},
  booktitle={Proceedings of the International AAAI Conference on Web and Social Media},
  volume={16},
  pages={536--547},
  year={2022}
}

@article{oddny2023impact,
  title={Impact of Reddit Community Culture on User Attitude Expression and Social Interaction},
  author={Oddn{\`y}, Lutgardis and Ainslie, Cecilia and Lakshman, Solly and Nathan, Dina},
  journal={Journal of Linguistics and Communication Studies},
  volume={2},
  number={4},
  pages={61--67},
  year={2023}
}

@article{razis2020modeling,
  title={Modeling influence with semantics in social networks: A survey},
  author={Razis, Gerasimos and Anagnostopoulos, Ioannis and Zeadally, Sherali},
  journal={ACM Computing Surveys (CSUR)},
  volume={53},
  number={1},
  pages={1--38},
  year={2020},
  publisher={ACM New York, NY, USA}
}


\section*{Appendix}
\appendix

\section{Annotation of Causal Language Detection}
We spent \$72 in our annotation task, which included \$18 in service fees to Prolific. Annotators, each assigned 25 comparisons to annotate via a Qualtrics survey link provided through Prolific, were paid \$4.5 each, with an hourly wage of \$27. We obtained Institutional Review Board (IRB) approval from our university. Prolific automatically deidentified participant data.

\begin{figure}[tbhp]
\centering
\includegraphics[width=0.9\linewidth]{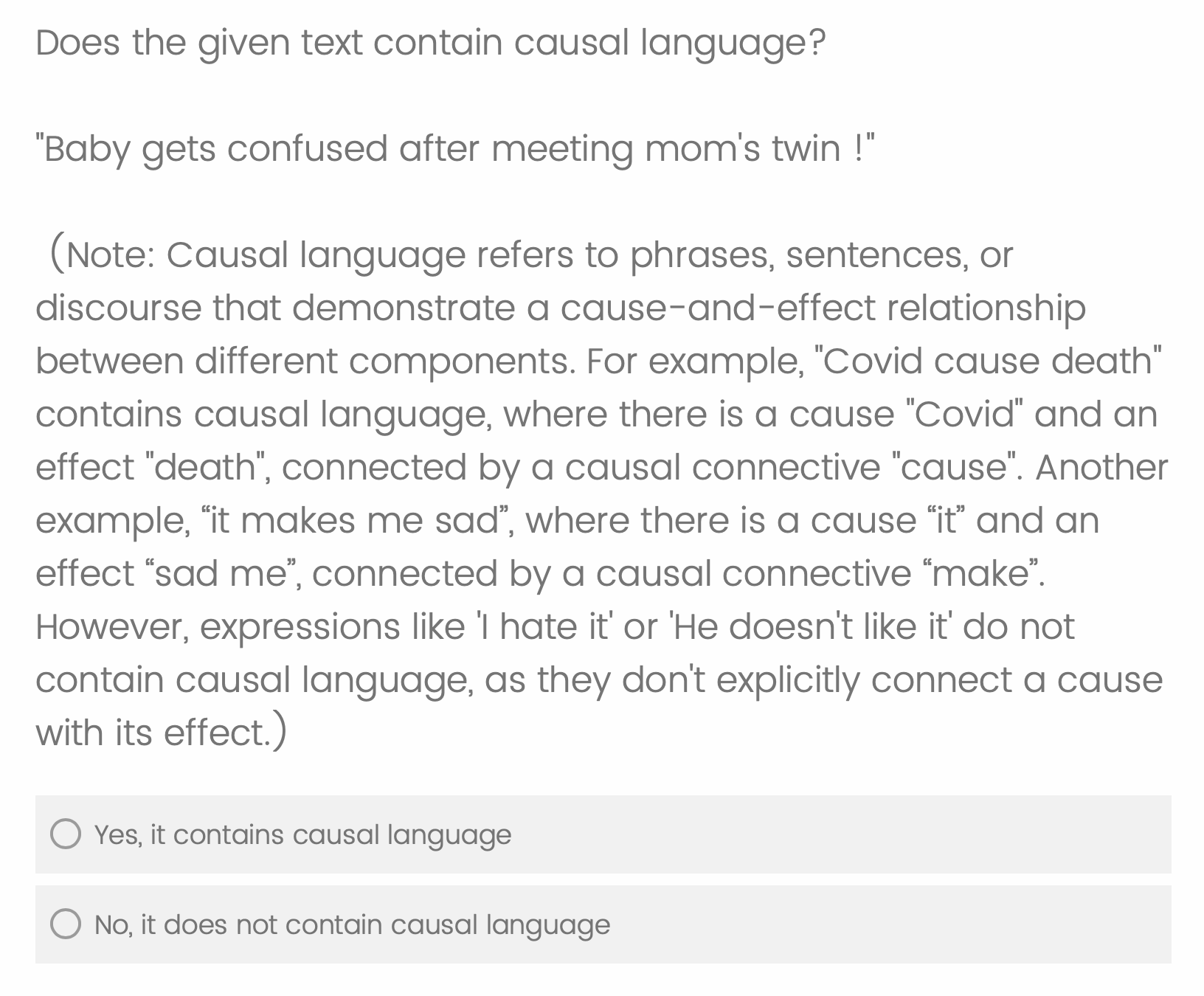}

\caption{Annotation example of causal language detection. This figure illustrates the human annotation task used to validate the causal language classifier described in Section 5 . Annotators were asked to determine whether a given post title contains a causal relationship, following training with labeled examples to ensure task understanding. The example highlights how annotators interpret causal structure in natural language rather than surface keywords alone, providing a qualitative check on model predictions.}
\label{fig:anno_CL}
\end{figure}

\section{Annotation of Matched Pairs from Counterfactual Experiment}
We spent \$20 in our annotation task, which included \$5 in service fees to Prolific. Annotators, each assigned 25 comparisons to annotate via a Qualtrics survey link provided through Prolific, were paid \$2.5 each, with an hourly wage of \$30. We obtained Institutional Review Board (IRB) approval from our university. Prolific automatically deidentified participant data.

\begin{figure}[tbhp]
\centering
\includegraphics[width=0.9\linewidth]{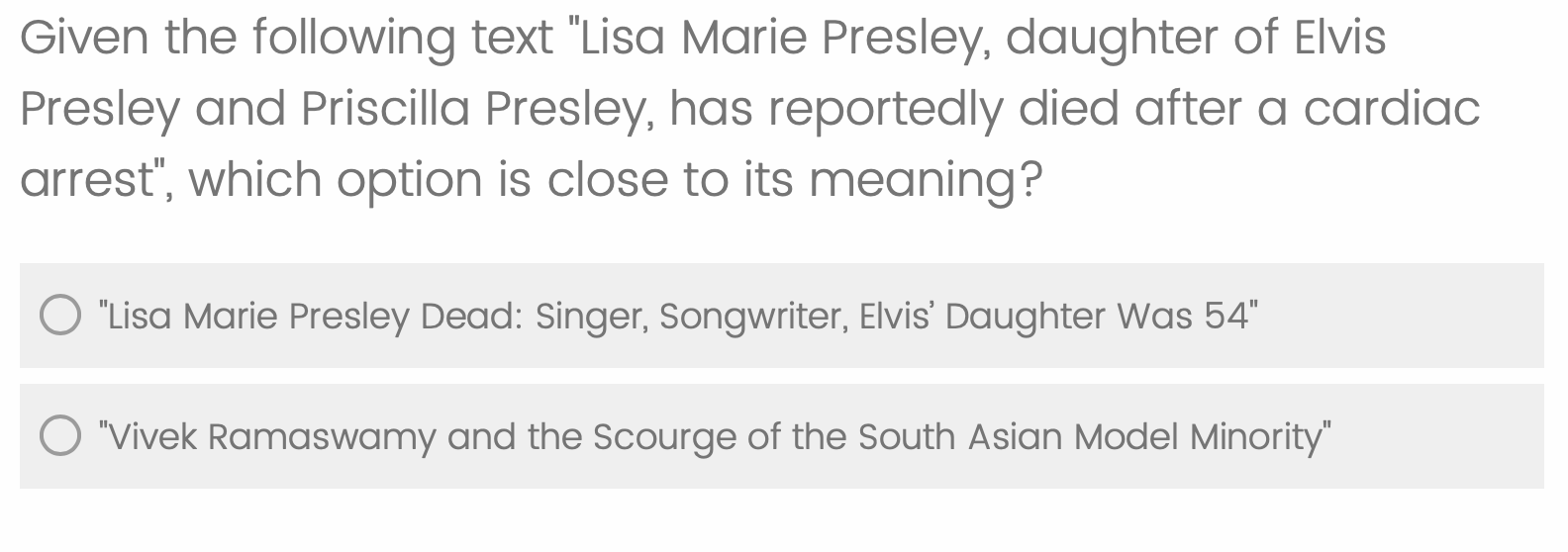}

\caption{Annotation example of matched pairs from counterfactual experiment. This figure shows the annotation interface used to validate the quality of counterfactual matching between causal and non-causal titles that conducted in Scetion 6. Annotators were presented with a causal title and two non-causal candidates (one from the counterfactual experiment and one from the baseline group) and asked which is semantically closer in meaning to the causal title. Please see Scetion 6.4 for validation details.}
\label{fig:anno_CF}
\end{figure}

\section{Figures}

\begin{figure}[tbhp]
\centering

\begin{subfigure}[b]{0.5\linewidth}
    \centering
    \includegraphics[width=\linewidth]{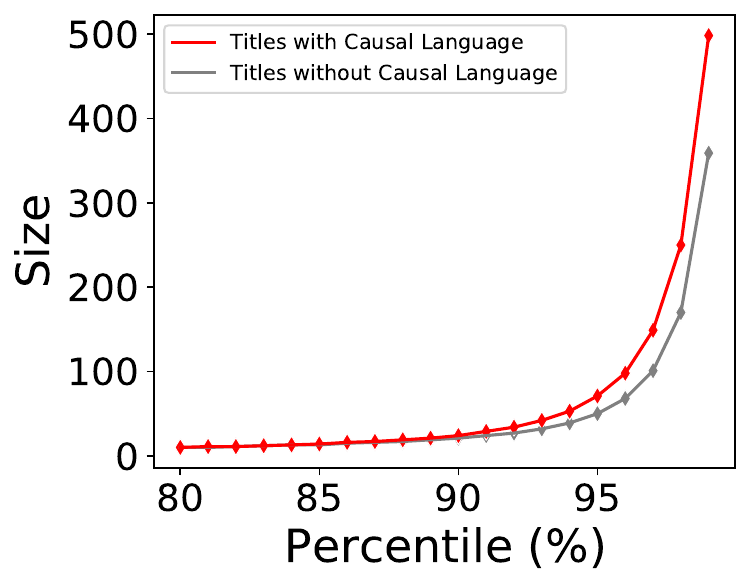}
    \caption{Size}
    \label{fig:RQ1_all_144_all_size}
\end{subfigure}

\vspace{1em} 

\begin{subfigure}[b]{0.48\linewidth}
    \centering
    \includegraphics[width=\linewidth]{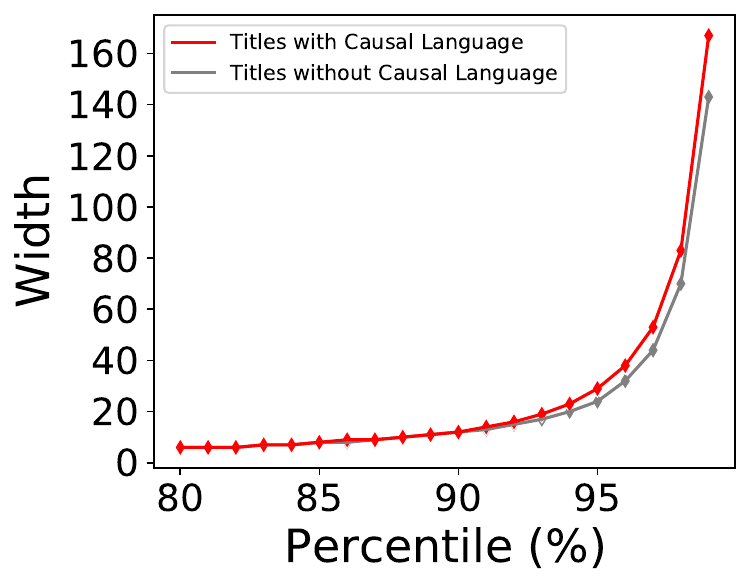}
    \caption{Width}
    \label{fig:RQ1_all_144_all_width}
\end{subfigure}
\hfill
\begin{subfigure}[b]{0.48\linewidth}
    \centering
    \includegraphics[width=\linewidth]{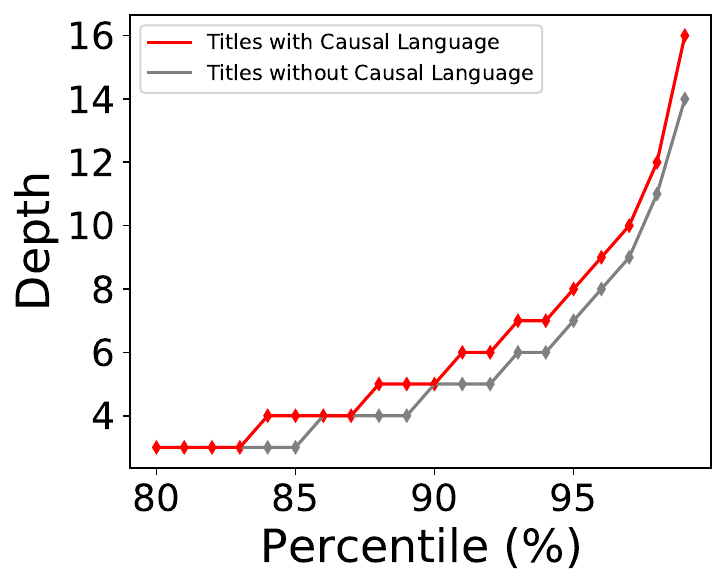}
    \caption{Depth}
    \label{fig:RQ1_all_144_all_depth}
\end{subfigure}

\caption{Comparison of post title \textbf{size},  \textbf{width} and  \textbf{depth} distributions between causal and non-causal language, across all 16,075,455 posts from 144 subreddits.}
\label{fig:RQ1_all_144_all}
\end{figure}

\end{document}